\documentclass[10pt,prd,aps,twocolumn,preprintnumbers, showpacs,nofootinbib,superscriptaddress,notitlepage]{revtex4-1}
\usepackage{amssymb,amsthm,amsmath}
\usepackage{graphicx}   % figures
\usepackage{color}      % color is used in text
\usepackage{slashed}    % Feynman slash
\usepackage{verbatim}
\usepackage{subcaption}
\usepackage[normalem]{ulem}
\usepackage{rotating}   % to rotate tables
\usepackage{multirow}   % multicolumn and multirow
\allowdisplaybreaks
\begin{document}
\title{A    Lattice QCD study of $p-\Lambda$ scattering  in   continuum  and chiral limits }

\author{Hang Liu}
\affiliation{Department of Physics, Shanghai Normal University, Shanghai 200234, China}

\author{Liuming Liu}
\affiliation{Institute of Modern Physics, Chinese Academy of Sciences, Lanzhou, Gansu Province 730000, China}

\author{Jin-Xin Tan}
\affiliation{INPAC, Key Laboratory for Particle Astrophysics and Cosmology (MOE),  Shanghai Key Laboratory for Particle Physics and Cosmology, School of Physics and Astronomy, Shanghai Jiao Tong University, Shanghai 200240, China}

\author{Wei Wang}
\affiliation{INPAC, Key Laboratory for Particle Astrophysics and Cosmology (MOE),  Shanghai Key Laboratory for Particle Physics and Cosmology, School of Physics and Astronomy, Shanghai Jiao Tong University, Shanghai 200240, China}
\affiliation{Southern Center for Nuclear-Science Theory (SCNT), Institute of Modern Physics, Chinese Academy of Sciences, Huizhou 516000, Guangdong Province, China}

\author{Haobo Yan}
\affiliation{School of Physics, Peking University, Beijing 100871, China}

\author{Qian-Teng Zhu}
\affiliation{INPAC, Key Laboratory for Particle Astrophysics and Cosmology (MOE),  Shanghai Key Laboratory for Particle Physics and Cosmology, School of Physics and Astronomy, Shanghai Jiao Tong University, Shanghai 200240, China}

\begin{abstract}
We present a first systematic study of $I=1/2$ proton-$\Lambda$ 
($p-\Lambda$) scattering from lattice QCD, using seven sets of $(2+1)$-flavor lattice ensembles with pion masses spanning 135–317 MeV and three lattice spacings with $a=(0.052,0.077, 0.105)$ fm. Using L\"uscher's finite-volume method,   effective range expansion and chiral/continuum extrapolations,  we obtain the inverse of scattering length and effective range for the $^1S_0$ channel as 0.177(83) GeV and 2.9(1.4) fm, and for the $^3S_1$ channel as 0.016(76) GeV and 1.8(1.1) fm. From the derived S-wave phase shifts, we provide an estimate of the $p-\Lambda$ scattering cross section.  Our results for scattering length, effective range and cross sections are in good agreement with available experimental measurements.  We also find that the $p-\Lambda$ system sustains attractive interactions. 
These results provide critical input for the unification of nuclear force theories and the construction of neutron star equations of state.

\end{abstract}

\maketitle

{\it Introduction: } Nucleon-$\Lambda$ scattering serves as an indispensable probe of strangeness $S=-1$ baryon-baryon dynamics, bridging the non-strange (nucleon) and strange ($\Lambda$ hyperon) sectors to address foundational questions across nuclear physics, astrophysics, and QCD. Its importance extends from constraining the $\Lambda$ single-particle potential in hypernuclei~\cite{Polinder:2006zh}—critical for unraveling the role of strangeness in nuclear matter—to resolving the ``hyperon puzzle" in neutron stars~\cite{Tolos:2020aln,Vidana:2018bdi,Vidana:2013nxa}: the tension between predicted hyperon-induced softening of the equation of state (EoS) and observations of massive neutron stars  $M>1.97M_{\odot}$  \cite{LIGOScientific:2018cki, Lonardoni:2014bwa}. Furthurmore, $p-\Lambda$ scattering data is critical for developing a unified theory of nuclear forces \cite{Gal:2016boi}, as it tests flavor symmetry breaking and non-perturbative QCD effects that elude direct computation.

On the experimental front, while early measurements of $p-\Lambda$ scattering cross sections date back to 1960s~\cite{Alexander:1968acu},  including a range of investigations in both low- and high-energy regimes~\cite{Kadyk:1971tc,Sechi-Zorn:1968mao,Hauptman:1977hr,Bassano:1967kbh,Charlton:1970bv,Anderson:1975rh,Mount:1975pb}, several important progress have made recently. CLAS collaboration reported a high-precision measurement of the $p\Lambda\to p\Lambda$ cross section between 0.9 and 2.0 GeV/c based on modern technologies~\cite{CLAS:2021gur}, while BESIII collaboration has used the beam pipe that contains the proton and the $\Lambda$ from $J/\psi\to \Lambda\bar \Lambda$  and reported their measurement of the $p\Lambda\to p\Lambda$ scattering at a $\Lambda$ momentum of 1.074 GeV/c~\cite{BESIII:2024geh}.  Using the data on two-particle correlations in  $\sqrt{s_{NN}}=3 $GeV Au+Au collisions from STAR,  Ref.~\cite{Hu:2023iib} has presented preliminary results on scattering length and effective range of $p-\Lambda$. Altogether, these recent high-precision results lay a solid experimental foundation for quantitative and precision studies of the $p−\Lambda$ interaction system.

From the theoretical aspect, understanding $p-\Lambda$ interactions has been hindered by the long-standing limitations in theoretical calculations. Phenomenological models, such as meson-exchange potentials~\cite{Rijken:1998yy,Holzenkamp:1989tq,Reuber:1992dh,Haidenbauer:2005zh,Nagels:1977ze,Maessen:1989sx,Rijken:2010zzb,Tominaga:2001ra}, constituent quark models~\cite{Zhang:1997ny,Fujiwara:2006yh} and Bethe-Salpeter approach~\cite{Chen:2025qxh}, relied on adjustable parameters, leading to large uncertain predictions for in key observables  like scattering lengths and effective ranges. While chiral effective field theory ($\chi$EFT) provides a systematic approach\cite{Polinder:2006zh,Haidenbauer:2013oca,Li:2016paq,Li:2016mln,Haidenbauer:2019boi,Song:2021yab,Haidenbauer:2023qhf}, its predictions depend on low-energy constants (LECs) calibrated to imprecise experimental data \cite{CLAS:2021gur, J-PARCE40:2021qxa}, resulting in less-precise constraints on the $p-\Lambda$ potential strength.  Early lattice QCD studies,  were constrained by quenched approximations \cite{Muroya:2004fz} or limited to unphysical pion masses ($m_\pi\gg 135$ MeV) on a coarse  lattice \cite{Beane:2006gf,Ishii:2006ec,Ishii:2012ssm,Beane:2010em,Nemura:2017bbw,Nemura:2017vjc,Nemura:2018tay}, making extrapolation to the physical point prone to systematic errors.  These theoretical ambiguities have propagated to downstream applications: models of neutron star EoS remain fragmented \cite{Ye:2024meg}, and the role of three-body forces in hypernuclei remains unresolvable \cite{Chatterjee:2015pua}.

To overcome these barriers, a precise lattice QCD calculation at the physical pion mass and in the continuum limit is imperative. In response to this demand, in this work, we present a systematic lattice QCD study of the $p-\Lambda$ interaction using the L\"uscher finite-volume method, which relates the energy spectrum in a finite volume to infinite-volume scattering parameters. Our calculations employ seven ensembles with three lattice spacings and various pion masses ranging from 135MeV to 317MeV, enabling a controlled extrapolation to the physical pion mass and continuum limit.  For the first time, we report the scattering lengths and effective ranges for both the  $^1S_0$ and $^3S_1$ channels at the physical pion mass and continuum limit, results that can be directly compared with experimental measurements. These results do not merely supplement the existing literature, they fill the critical gap in hyperon-nulceon interaction data, providing useful inputs to help resolve the hyperon puzzle, advance the unified theory of nuclear forces, and quantify QCD flavor symmetry breaking.

{\it Numerical setup: } Calculation presented in this Letter is based on the gauge configurations generated by the CLQCD collaboration with 2+1 dynamical quark flavors, employing the tadpole-improved Symanzik gauge action and Clover fermion action~\cite{CLQCD:2023sdb}. We use seven ensembles with three lattice spacing values: $a$ = 0.105~fm, 0.077~fm and 0.052~fm, and various pion masses ranging from 135~MeV to 317~MeV. 
The diverse parameters across these ensembles facilitate a robust extrapolation of our results to the physical pion mass and continuum limit. 
The detailed information regarding these ensembles is provided in Table~\ref{tab:conf}.

\begin{table}[h]
\begin{tabular}{c|cccc}
\hline
   ID &  $a\rm{(fm)}$ & $(L/a)^3\times T/a$   & $m_{\pi}\rm{(MeV)}$& $N_{\rm{conf}}$ \\\hline
 C24P29 &$0.10530$ & $24^3 \times 72$   &$292.7(1.2)$ & $872$\\
 C32P29 &$0.10530$ & $32^3 \times 64$   &$292.4(1.1)$ & $984$\\
 C48P23 &$0.10530$ & $48^3 \times 96$   &$225.6(0.9)$ & $265$\\
 C48P14 &$0.10530$ & $48^3 \times 96$   &$135.5(1.6)$ & $259$\\
F48P21 &$0.07746$ & $48^3 \times 96$   &$207.2(1.1)$ & $220$\\
 F48P30 &$0.07746$ & $48^3 \times 96$   &$303.4(0.9)$ & $359$\\
 H48P32 &$0.05187$ & $48^3 \times 144$   &$317.2(0.9)$ & $274$\\
\hline
\end{tabular}
\caption{ Parameters of lattice ensembles: the lattice spacing $a$, lattice size $(L/a)^3\times T/a$, the mass of pion $m_{\pi}$ and the number of configurations $N_{\rm{conf}}$. }
\label{tab:conf}
\end{table}

Quark propagators are computed using the distillation quark smearing method~\cite{HadronSpectrum:2009krc}. The smearing operator is composed of a small number ($N_{ev}$) of the eigenvectors associated with the $N_{ev}$ lowest eigenvalues of the three-dimensional Laplacian defined in terms of the HYP-smeared gauge fields. The number of eigenvectors $N_{ev}$ is 100 for the ensembles with $L/a$ = 24 and 32, and 200 for the ensembles with $L/a$=48. Further increases in $N_{ev}$ do not alter the extracted spectrum~\cite{Chen:2025xww}.

{\it Finite-volume spectrum: } 
To obtain the finite-volume spectrum of the $p-\Lambda$ system, we start with the construction of the interpolating operators. In continuum space, the rotational symmetry is described by the group $SU(2)$, its irreducible representations(irreps) are labeled by the total angular momentum $J=0,1/2,1,\cdots$. While on the lattice, the rotational symmetry is reduced to the double cover of the cubic group, $O^D$. Including parity extends this group to $O_h^D = O^D \otimes Z_2$, with $Z_2$ accounting for the spatial inversion. In moving frames, the symmetry is further reduced to little groups, and operators must be constructed in the irreps of $O_h^D$ and its little groups. 

In this work, we study the S-wave $p-\Lambda$ scattering in both $^1S_0$ and $^3S_1$ channels, and construct the operators in the rest frame as well as the moving frame with total momentum $\vec{P}=(0,0,1)$ (in units of $\frac{2\pi}{L}$). For the $^1S_0$ channel, the relevant irreps are $A_1^+$ in the rest frame and $A_1$ in the moving frame. For the  $^3S_1$ channel, the relevant irreps are $T_1^+$ in the rest frame and $E_2$ in the moving frame. The general form of the $p-\Lambda$ operators can be written as
\begin{equation}
O^{\Lambda, \lambda, \vec{P}}_{p_1, p_2} = \sum_{\mu, \nu, \vec{p_1}, \vec{p_2}}c^{\Lambda, \lambda, \vec{P}}_{\mu, \nu, \vec{p_1}, \vec{p_2}} p_\mu(\vec{p_1}) \Lambda_\nu(\vec{p2}), \vec{P} = \vec{p_1} + \vec{p_2}.
\end{equation}
where $p$ and $\Lambda$ denote the single-particle operators for proton and $\Lambda$, respectively, and $\mu$, $\nu$ are spinor indices. $\Lambda$ labels the irrep and $\lambda$ denotes its row. The sum runs over all momenta $\vec{p_1}(\vec{p_2})$ with fixed magnitude $p_1(p_2)$ that are related by the rotations of the cubic group. The coefficients $c^{\Lambda, \lambda, \vec{P}}_{\mu, \nu, \vec{p_1}, \vec{p_2}}$ are chosen so that the operator transforms under the irrep $\Lambda$.  These coefficients are constructed following the method of Ref.~\cite{Yan:2025jlq} and are computed using the open-source package \textbf{OpTion}~\cite{Yan:2025jlq} provided therein. The single particle operators and the explicit form of the $p-\Lambda$ two particle operators are provided in the supplemental materials~\cite{supplemental}. 

To validate a single-channel two-body scattering analysis, we restrict the energy range below the $p-\Sigma$ threshold and the open three-body thresholds (e.g., $p\Lambda\pi$). Accordingly, we use one operator with $p_1 = p_2 =0$ in the rest frame and two operators in the moving frame $\vec{P}=(0,0,1)$: one with $p_1=1, p_2 =0$ and the other one with $p_1 = 0, p_2 =1$. 

The spectrum is extracted from correlation functions of the operators introduced above:
\begin{eqnarray}
C_{ij}(t)=\sum_{t_s} \langle O_i(t+t_s)O_j^\dagger(t_s)\rangle,
\end{eqnarray}
where $i,j$ index the two-particle operators. In the rest frame, one operator is used. For the moving frame, $C$ is a $2\times2$ matrix. The source time $t_s$ runs over all time slices to increase statistics. 

Solving the generalized eigenvalue problem(GEVP)~\cite{Luscher:1990ck}:
\begin{eqnarray}
C(t) v_n(t,t_0) = \lambda_n(t, t_0) C(t_0)v_n(t,t_0), 
\end{eqnarray}
the energies can be obtained by fitting the eigenvalues $\lambda_n(t, t_0)$ to an exponential form. In practice, we employ the following ratio to subtract the non-interacting energy:
\begin{eqnarray}
    \mathcal{R}_n(t,t_0)&=&\frac{\lambda_n(t,t_0)}{C_p(t-t_0)C_{\Lambda}(t-t_0)}\notag\\
   & \xrightarrow{t\to \infty}& A_n e^{-\Delta E_n (t-t_0)},
\end{eqnarray}
where $C_p$ and $C_{\Lambda}$ are the single-particle correlation functions of the $p$ and $\Lambda$ at zero momentum, respectively, and $\Delta E_n$ is the energy shift relative to  threshold:
\begin{eqnarray}
    \Delta E_n=E_n-m_{p}-m_{\Lambda}.
\end{eqnarray}
The energy shift $\Delta E_n$  is extracted by fitting $\mathcal{R}_n(t,t_0)$ to the form $ A_n e^{-\Delta E_n (t-t_0)}$ at a time window where the effective mass exhibits a clear plateau. All fits yield $\chi^2$ per degree of freedom around or less than one. The final results for $\Delta E_n$ are summarized in Tab.~\ref{tab: delta_E}. More fitting details are provided in the supplemental materials~\cite{supplemental}. 
\begin{table}[h]
    \centering
    \begin{tabular}{|c|c|c|cc|cc|}
    \hline
    &\multicolumn{2}{c|}{$\vec{P}=(0,0,0)$}&\multicolumn{4}{c|}{$\vec{P}=(0,0,1)$}\\
    \hline
    &$A^+_1$&$T^+_1$&\multicolumn{2}{c|}{$A_1$}&\multicolumn{2}{c|}{$E_2$}\\
    \hline
    $\Delta E(\rm MeV)$&$\Delta E_0$&$\Delta E_0$&$\Delta E_1$&$\Delta E_2$&$\Delta E_1$&$\Delta E_2$\\
    \hline
     C24P29&$-28(7)$&$-21(5)$&&&&\\
    \hline
    C32P29 &$-19(2)$&$-9(3)$& $39(3)$&$49(3)$&$47(1)$&\\
    \hline
    C48P23 &$-4(1)$&$-3(1)$& $21(4)$&$22(2)$&$18(1)$&$22(3)$\\
    \hline
    C48P14 &$-6(2)$&$-3(2)$& $-6(4)$&$18(6)$&$8(3)$&$15(2)$\\
    \hline
    F48P30 &$-6.7(8)$&$-6(1)$& $31(2)$&$40(3)$&$33(3)$&$44(4)$\\
    \hline
    F48P21 &$-7(2)$&$-9(2)$& $34(6)$&$60(7)$&$31(6)$&$40(7)$\\
    \hline
    H48P32 &$-12(3)$&$-10(2)$& $69(11)$&$109(18)$&$94(10)$&$123(9)$\\
    \hline
    \end{tabular}
    \caption{Results for the energy difference $\Delta E$ in unit $\rm MeV$ in the center-of-mass frame $\vec{P}=(0,0,0)$ and the moving frame $\vec{P} = (0,0,1)$ obtained in our simulations for the ensembles. Note: $\Delta E_n= E_n- (m_p+m_\Lambda)$. Empty entries indicate no reliable extraction due to statistical noise or excited-state contamination.}
    \label{tab: delta_E}
\end{table}

{\it Scattering analysis:} Scattering parameters are determined using L\"uscher's finite-volume method~\cite{Luscher:1986pf, Luscher:1990ux, Luscher:1990ck}, which connects the finite-volume spectrum with the scattering phase shift in infinite volume. Ignoring the effects of the higher partial wave mixing, L\"uscher's formula in all the irreps considered here reduces to:
\begin{eqnarray}
%q \cot{\delta_0(q)} = \frac{1}{\pi^{3/2} \gamma} \mathcal{Z}_{00}^{\mathbf{\Delta}(m_1,m_2)}(1; q^2),
q \cot{\delta_0(q)} = \frac{1}{\pi^{3/2} \gamma} \mathcal{Z}_{00}^{\bf{d}}(1; q^2),
\label{eq:Luscher}
\end{eqnarray}
where $\mathcal{Z}_{00}^{\bf{d}}(1; q^2)$ is the generalized zeta function that can be evaluated numerically for a given $q$. The dimensionless variable $q$ is defined in terms of the scattering momentum $k$ as $q = kL/2\pi$, and $k$ is related to the center-of-momentum frame energy $E^* = \sqrt{m_1^2 + k^2} + \sqrt{m_2^2 + k^2}$, where $m_{1,2}$ are the masses of the two scattering particles. In a moving frame, the zeta function also depends on the Lorentz factor $\gamma = E/E^*$, with $E$ denoting the energy in the moving frame, and the coefficient $A \equiv 1 + \frac{m_1^2 - m_2^2}{E^{*2}}$, which accounts for the mass difference of the two particles~\cite{Leskovec:2012gb}. The detailed expression for the zeta function is provided in the supplemental materials~\cite{supplemental}. 

Near the threshold, the phase shift can be parametrized using the effective range expansion:
\begin{eqnarray}
    k^{2l+1}\cot \delta_{l}(k)=\frac{1}{a_l}+\frac{1}{2}r_l k^2+..., 
\label{eq:ERE}
\end{eqnarray}
where $a_l$ is the scattering length and $r_l$ is the effective range for partial wave $l$, and the ellipsis represents terms that are higher order in $k^2$. 
 
In this work, we focus on $S$-wave($l=0$) contributions and neglect the mixing effects from $D$-waves and higher partial waves ($l \geq 2$), which are expected to give only minor corrections to the $S$-wave observables.  We compute the values of $k\cot\delta_0$ using Eq.~\ref{eq:Luscher} and obtain the S-wave scattering length $a_0$ and effective range $r_0$ by fitting the data to Eq.~\eqref{eq:ERE}. 

\begin{figure}[h]
\centering
\includegraphics[width=0.48\textwidth]{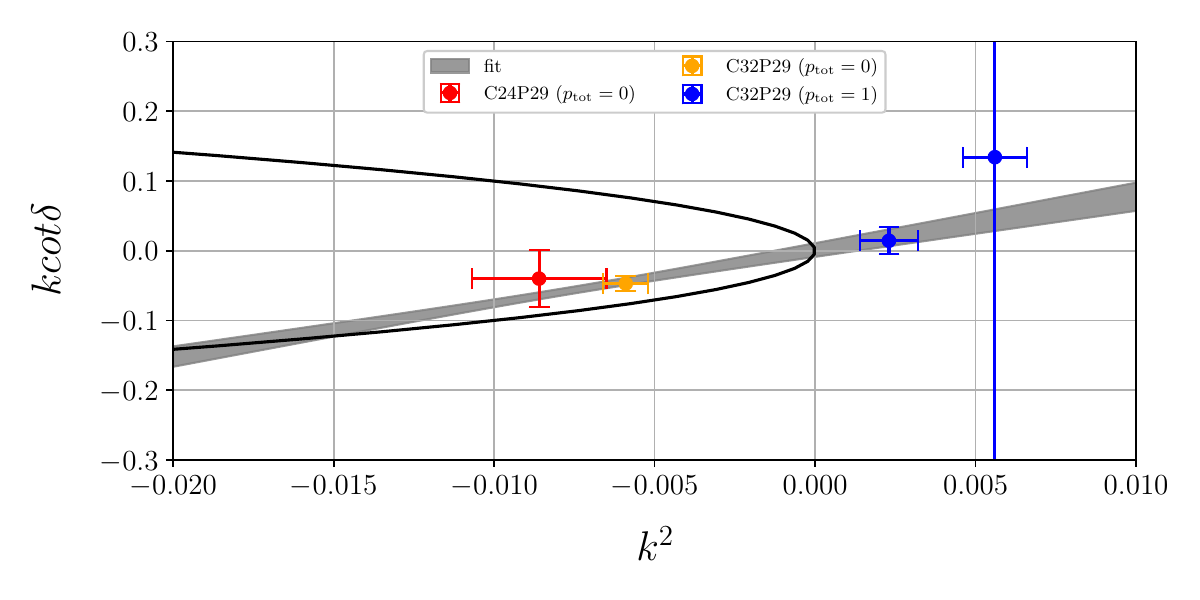} 
\includegraphics[width=0.48\textwidth]{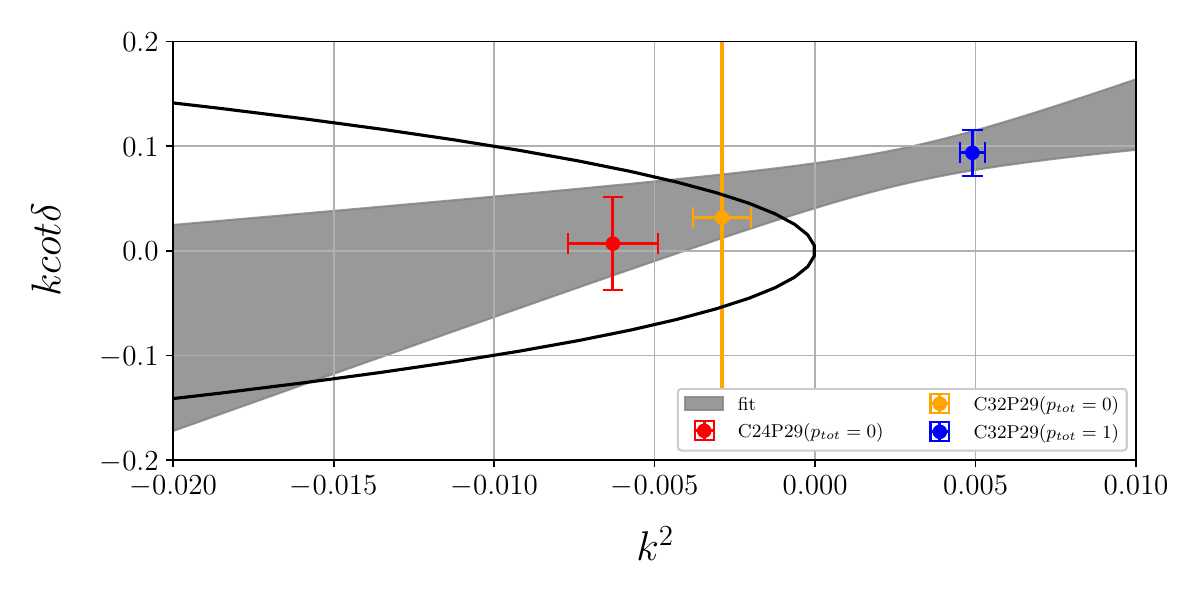} 
  \caption{The values of $k\cot\delta$ as a function of $k^2$. The data points are from the ensembles C24P29 and C32P29, having the same pion mass and lattice spacing. The gray bands are the fits to Eq.~\eqref{eq:ERE} and the width of the bands indicate 1$\sigma$ statistical uncertainty. The black curves are $ik$. The upper and lower panels are for the $^1S_0$ and $^3S_1$ channels, respectively.}
\label{fig:ERE_fit_C24-32-A1_T1}
\end{figure}

As an example, Fig.\ref{fig:ERE_fit_C24-32-A1_T1} shows the fit for the ensembles with $m_\pi = 292$MeV and lattice spacing $a=0.1053$fm. Fits for the remaining ensembles are provided in the Supplemental Material~\cite{supplemental}. The extracted scattering lengths and effective ranges for all ensembles are summarized in Table~\ref{tab:a0-r0-A1-T1}.

\begin{table}[h]
    \centering
    \begin{tabular}{c|c|c|c|c}
    \hline
    &\multicolumn{2}{c|}{$^1S_0$}&\multicolumn{2}{c}{$^3S_1$}\\
    \hline
    &$a_0^{-1}(\rm GeV)$&$r_0(\rm fm)$&$a_0^{-1} (\rm GeV)$&$r_0(\rm fm)$\\
    \hline
{\scriptsize C24P29/C32P29}&$0.0002^{+0.0195}_{-0.0156}$&$1.58^{+0.24}_{-0.19}$&$0.110^{+0.034}_{-0.025}$&$1.61^{+0.45}_{-0.74}$\\
\hline
C48P23&$0.077^{+0.056}_{-0.034}$&$1.74^{+2.19}_{-2.85}$&$0.094^{+0.139}_{-0.060}$&$-10.1^{+6.5}_{-18.8}$\\
    \hline    C48P14&$0.054^{+0.066}_{-0.034}$&$3.81^{+1.53}_{-0.82}$&$-0.028^{+0.005}_{-0.005}$&$2.89^{+0.55}_{-0.40}$\\
    \hline
    F48P30&$0.052^{+0.034}_{-0.032}$& $-1.60^{+1.13}_{-1.46}$&$0.106^{+0.053}_{-0.047}$&$-1.40^{+1.48}_{-2.05}$\\
    \hline
    F48P21&$0.061^{+0.086}_{-0.078}$ &$-1.75^{+2.47}_{-4.68}$&$0.017^{+0.059}_{-0.072}$&$-0.87^{+2.08}_{-3.55}$\\
    \hline    H48P32&$0.178^{+0.321}_{-0.176}$&$-1.32^{+3.15}_{-6.12}$&$0.201^{+0.190}_{-0.128}$&$-3.68^{+1.20}_{-2.26}$\\
    \hline
    \end{tabular}
 \caption{The fitting result for the $^1S_0$-wave and $^3S_1$-wave scattering using the ERE parametrization. ERE fit results for the C24P29 and C32P29 ensembles are listed in the first column. The distribution of the 3000 bootstrap replicates exhibits significant non-Gaussian asymmetry. We report the median as the central estimate, with uncertainties corresponding to the $1\sigma$ equivalent quantiles for  upper and lower bounds.
 }
\label{tab:a0-r0-A1-T1}
\end{table}

\begin{figure}[h]
\centering
\includegraphics[width=0.4\textwidth]{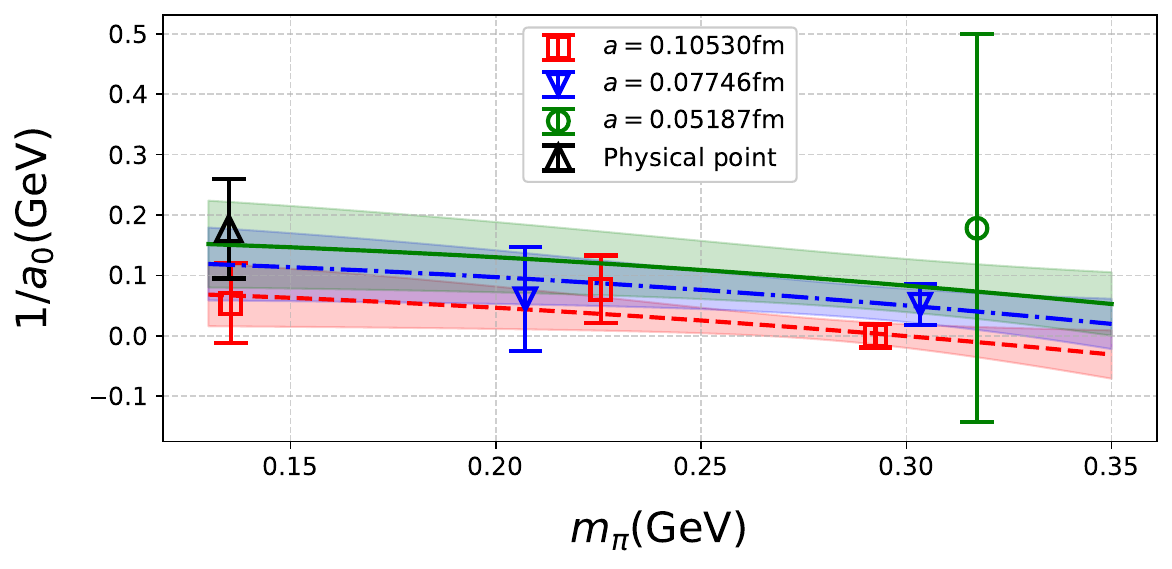}
\includegraphics[width=0.4\textwidth]{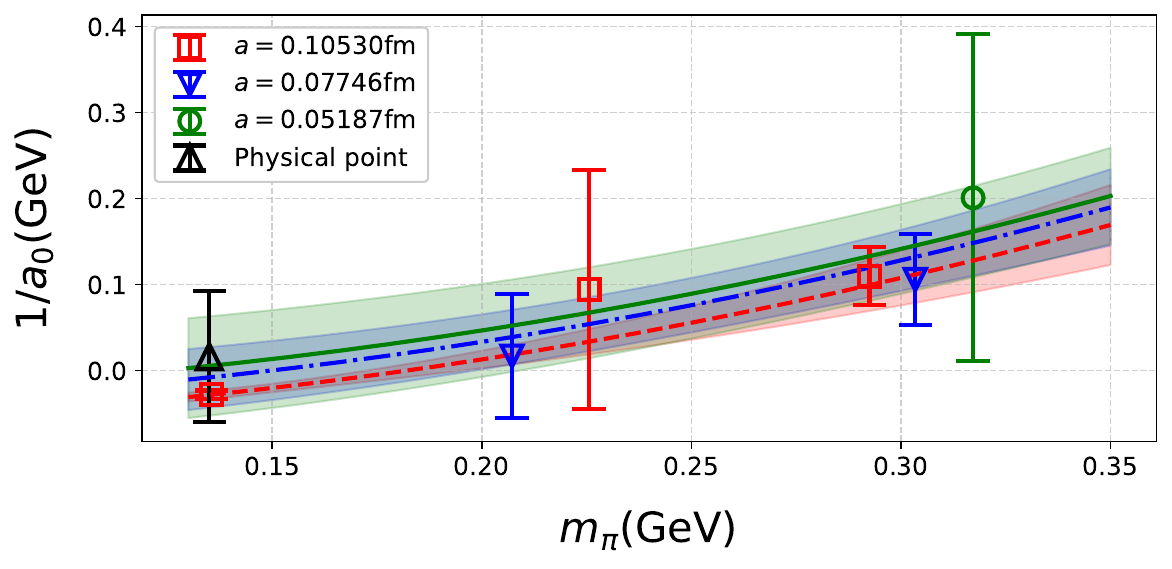}
\caption{Results for the $1/a_0$ extraplation according to Eq.~\eqref{eq:joint-extrapolate} for the channel $^1S_0$ (upper panel) and $^3S_1$ (lower panel) with the corresponding $\chi^2/d.o.f=0.15$ and  $\chi^2/d.o.f=0.18$. The black point corresponds to the physical point and continuum limit.} 
\label{fig:a0_extraplation}
\end{figure}

\begin{figure}[h]
\centering
\includegraphics[width=0.4\textwidth]{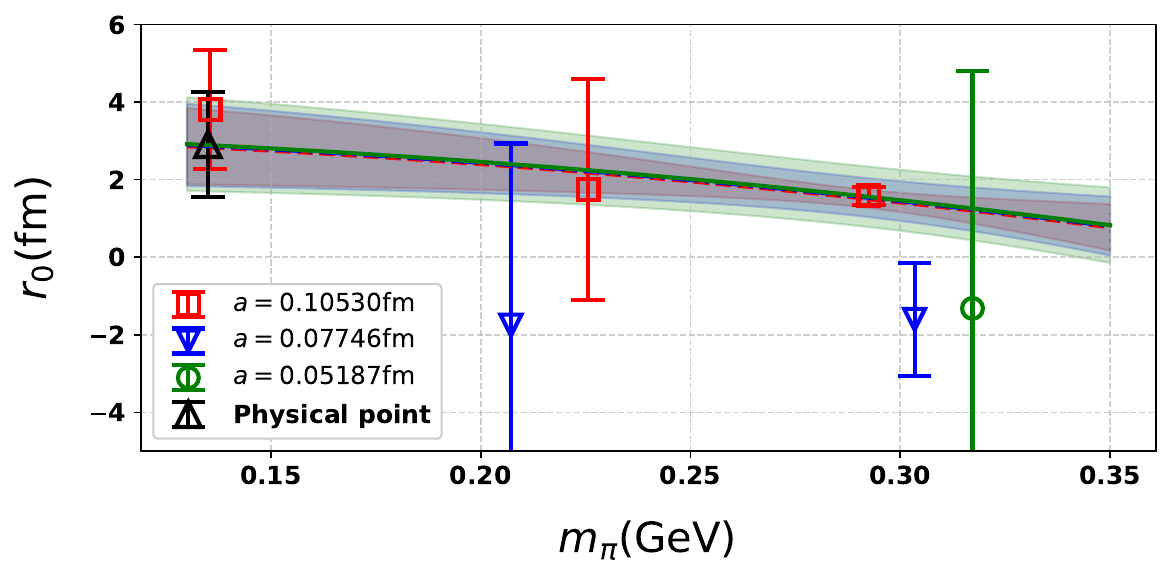}
\includegraphics[width=0.4\textwidth]{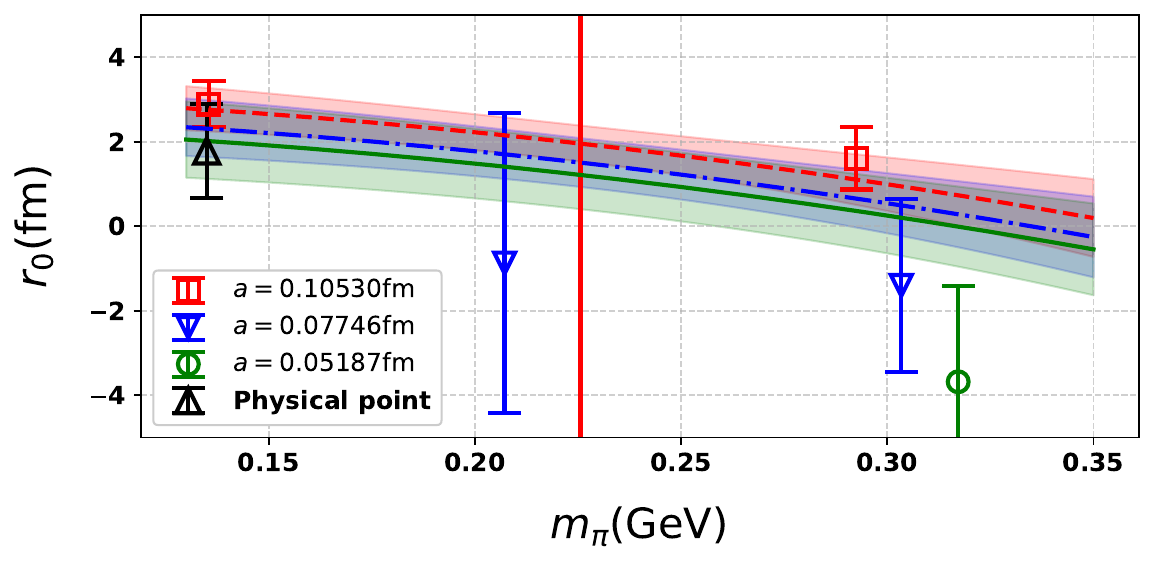}
\caption{Results for the $r_0$ extraplation according to Eq.~\eqref{eq:joint-extrapolate} for the channel $^1S_0$ and $^3S_1$  with the corresponding $\chi^2/d.o.f=1.2$, and the corresponding $\chi^2/d.o.f=1.1$. The black point corresponds to the physical point and continuum limit. } 
\label{fig:r0_extraplation}
\end{figure}

{\it Chiral and Continuum extrapolation.}
For both $^1S_0$ and $^3S_1$ channels, we extrapolate the inverse of scattering length and the effective range to  
physical pion mass  and continuum limit using the following
parametrization~\cite{Yan:2024yuq}:  
\begin{eqnarray}
a_0^{-1}&=& a^{-1}_{0,\rm phys} + c_1(m_\pi^2 -m_{\pi, {\rm phys}}^2) + c_2 a^2,\notag\\
r_0 &=& r_{0,\rm phys} + c_3(m_\pi^2 -m_{\pi, {\rm phys}}^2) + c_4 a^2.
\label{eq:joint-extrapolate}
\end{eqnarray}
For the $^1S_0$ channel, the extrapolated results are 
\begin{eqnarray}
a^{-1}_{0,\rm phys}&=&0.177 (83) \,{\rm GeV},\;
r_{0,\rm phys}=2.9 (1.4) \,{\rm fm},
\end{eqnarray}
while  the $^3S_1$ channel, they are 
\begin{eqnarray}
a^{-1}_{0,\rm phys}&=&0.016 (76) \,{\rm GeV},\;
r_{0,\rm phys}=1.8 (1.1) \,{\rm fm}. 
\end{eqnarray}
The extrapolations of $1/a_0$ and $r_0$ for $^1S_0$ and $^3S_1$  are shown in Fig.~\ref{fig:a0_extraplation} and Fig.~\ref{fig:r0_extraplation}, respectively.

{\it Discussions:}
A critical benchmark for our lattice results is direct comparison with experimental data on 
scattering parameters.  
The STAR detector at RHIC has provided the spin-averaged experimental values:~\cite{Hu:2023iib}: 
\begin{eqnarray}
a_0&=&(2.32_{-0.11}^{+0.12}) \, {\rm fm},\;\;\;
r_0=(3.5_{-1.3}^{+2.7}) \, {\rm fm}. 
\end{eqnarray}
For our lattice results, we perform spin averaging of the  $^1S_0$ and $^3S_1$ channel parameters, yielding  
\begin{eqnarray}
a_{0,\rm avg}&=& 3.5(3.8) \, {\rm fm},\;\;\;
r_{0,\rm avg}=2.08(90) \, {\rm fm}.
\end{eqnarray} 
These values are consistent with the STAR measurements within statistical errors, marking the first direct agreement between lattice QCD predictions and experimental data for  $p-\Lambda$ scattering. 

\begin{figure}[h!]
\centering
\includegraphics[width=0.4\textwidth]{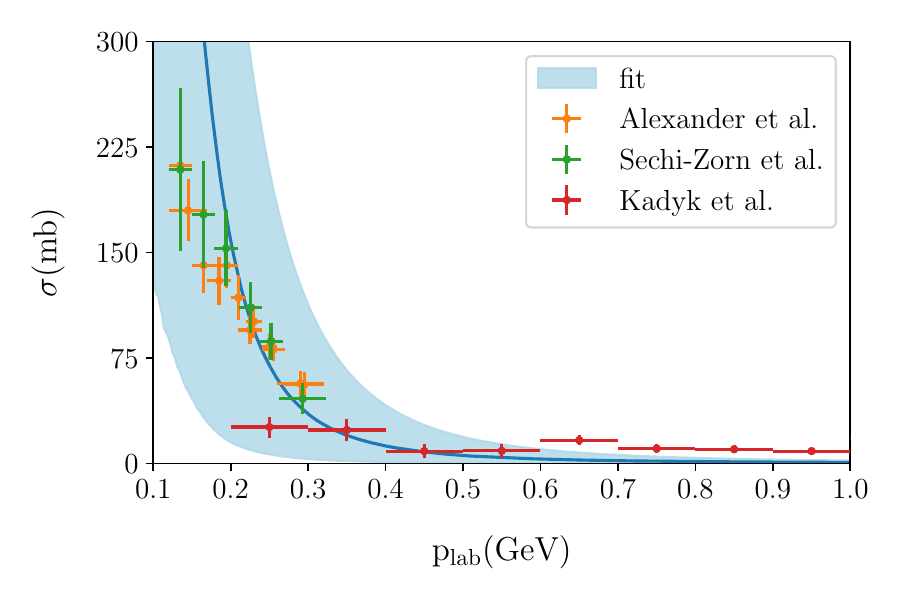}
\caption{Comparison of  the spin-averaged cross section with experimental meesurements~\cite{Alexander:1968acu,Sechi-Zorn:1968mao,Kadyk:1971tc}}. 
\label{fig:cross-section}
\end{figure}
Using the extracted 
s-wave phase shifts and the standard cross-section formula \begin{eqnarray}
\sigma_l=\frac{4\pi}{k^2}(2l+1)\sin^2\delta_l, 
\end{eqnarray}
we compute the spin-averaged s-wave $p-\Lambda$  scattering cross-section and compare it with available experimental data~\cite{Alexander:1968acu,Sechi-Zorn:1968mao,Kadyk:1971tc}. As shown in Fig.~\ref{fig:cross-section}, our results have a larger central value but still  align with experimental measurements within error bars for most momentum ranges. The slight discrepancy at low momenta ($p_{\rm lab}\leq 0.2$ GeV) may stem from two factors: first, the amplified statistical uncertainties near threshold (where small fluctuations in energy levels propagate to $k\cot\delta$ via L\"uscher’s relation); second, the limited applicability of the single-channel ERE in capturing subtle threshold behaviors of hyperon-nucleon interactions.

The $k \cot\delta - k^2$ plots presented in the previous section show a consistent pattern of attractive interactions across all simulated pion masses. A sufficiently strong attraction can generate poles in the analytically-continued scattering amplitude, which lives on multiple Riemann sheets (RSs) in the complex $s = E_{\mathrm{CM}}^2$ plane.  The scattering amplitude with partial wave $l$ and total angular momentum $J$ is proportional to  ${(k\cot\delta_l^{(J)}-ik)}^{-1}$~\cite{Luscher:1986pf}. The scattering t-matrix
can be written as
\begin{equation}
    t_l^{(J)} \sim \frac{\sqrt{s}}{2}\frac{1}{k\cot\delta_l^{(J)}-ik}.
\end{equation}
Positions of poles determine whether the system exhibits a bound state, a virtual state, or a resonance. In particular, the intersection of  ERE curve with upper half of the parabola defined by $ik$ indicates a pole below threshold on the real energy axis of the second RS, corresponding to a virtual state.

\begin{figure}[h]
\centering
\includegraphics[width=0.4\textwidth]{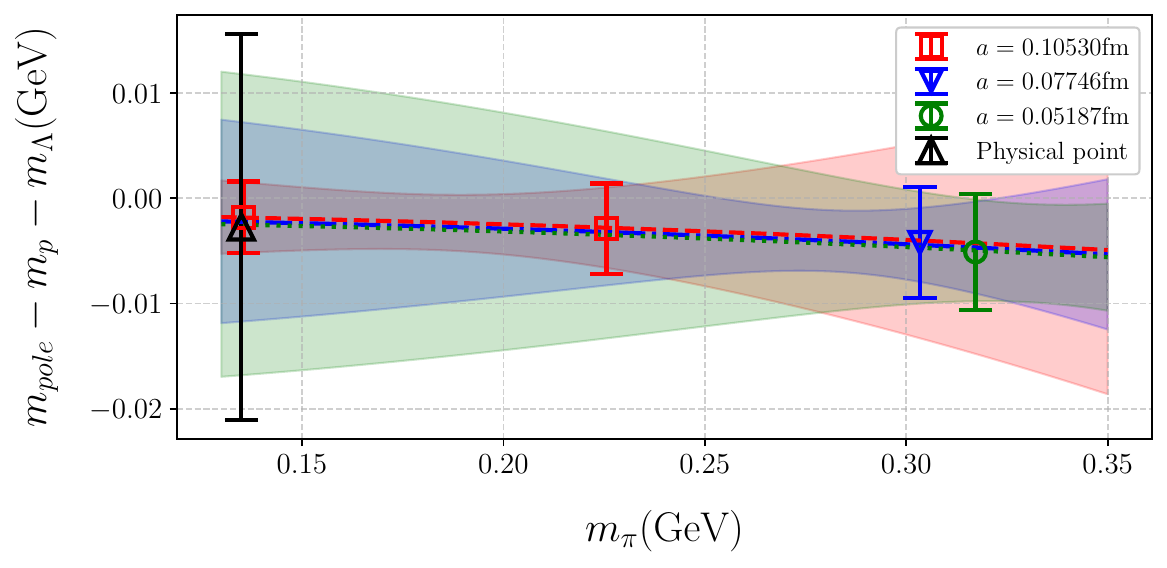} 
\caption{The $m_{pole}-m_p-m_{\Lambda}$ of the virtual state for $^1S_0$. We performed a joint extrapolation($m_{pole} = m_{pole,\rm phys} + c_5(m_\pi^2 -m_{\pi, {\rm phys}}^2) + c_6 a^2$) of the pole trajectory to the physical point. The mass difference at the physical point are $-3(18)\rm MeV$. } 
\label{fig:pole}
\end{figure}

For the channel $^1S_0$, the resulting pole positions extracted from all ensembles used in this work are summarized in Fig.~\ref{fig:pole}, yielding a $p-\Lambda$  mass difference of $-3(18)$ MeV after chiral and continuum extrapolation, consistent with the attractive nature of the $p-\Lambda$  interaction inferred from our scattering parameters. For the channel $^3S_1$, reducing the pion mass from 300 MeV toward its physical value induces a continuous migration of the $S$-matrix pole. The pole trajectory originates on the second Riemann sheet, approaches the threshold, and subsequently moves onto the physical sheet. This evolution suggests that the state, initially virtual at larger $m_\pi$, may transition into a bound-state-like pole as the physical point is approached.

{\it Summary.} We have presented a first systematic lattice QCD study of the $p-\Lambda$ scattering based on seven gauge ensembles with $2+1$ dynamical flavors at pion masses from $135.5 \,\rm MeV$ to $317.2 \,\rm MeV$ and of three different lattice spacings. We investigate the $p-\Lambda$ interaction through a systematic analysis of single- and two-particle correlation functions. Utilizing the effective range expansion and incorporating chiral and continuum extrapolations, we find that the $p-\Lambda$ system exhibits weak attractive interactions within the considered energy regime.

Extrapolating to the physical point and the continuum limit, we obtain the inverse of scattering length and effective range for the $^1S_0$ channel as $0.177(83)\ \text{GeV}$ and $2.9(1.4)\ \text{fm}$. For the $^3S_1$ channel, we find $1/a_0 = 0.016(76)\ \text{GeV}$ and $r_0 = 1.8(1.1)\ \text{fm}$. These results are consistent with spin-averaged experimental results.  

The next key objective in our study of hyperon–nucleon interactions is a full $p\Lambda$–$N\Sigma$ coupled-channel analysis~\cite{ALICE:2021njx}, as well as the extension to additional channels involving hyperons. The explicit inclusion of a physical ensemble with a smaller lattice spacing in the future may greatly reduce the systematic error.

{\it Acknowledgments.} 
We thank Lie-Wen Chen, Meng-Lin Du, Lisheng Geng, Feng-Kun Guo, Jun He,  Y. Hu, Chuan Liu, Tao Luo, Yu Lu, Yan Lyu, Yi-Feng Sun, Jia-Jun Wu, Hanyang Xing, Jun-Ting Ye, and Kang Yu for useful discussions.  We thank the CLQCD collaborations for providing us the gauge configurations with dynamical fermions~\cite{CLQCD:2023sdb}, which are generated on the HPC Cluster of ITP-CAS, the Southern Nuclear Science Computing Center(SNSC), the Siyuan-1 cluster supported by the Center for High Performance Computing at Shanghai Jiao Tong University, and the Dongjiang Yuan Intelligent Computing Center.  This work is supported in part by National Natural Science Foundation of China under grants No.12125503, 12305103, 12405101, and 124B2096.

\clearpage
\begin{appendix}

\begin{widetext}
\section{Baryon Operators, correlation functions and dispersion relations}
For the proton and $\Lambda$, both of which have quantum numbers $J^P = {\frac{1}{2}}^+$, the interpolating operators are:
\begin{eqnarray}
    \mathcal{O}^{\mu}_{p}(\vec{p},t)&=&\sum_{\vec{x}}\epsilon^{abc}[u^{aT}(\vec{x},t)(C\gamma_5)d^b(\vec{x},t)](u^c)^{\mu}(\vec{x},t)e^{i\vec{p}\cdot \vec{x}}\notag\\
    \mathcal{O}^{\mu}_{\Lambda}(\vec{p},t)&=&\sum_{\vec{x}}\epsilon^{abc}[u^{aT}(\vec{x},t)(C\gamma_5)d^b(\vec{x},t)](s^c)^{\mu}(\vec{x},t)e^{i\vec{p}\cdot \vec{x}}, 
\end{eqnarray}
where $\vec{p}$ and $\vec{x}$ are three-dimensional vectors. For convenient, one can define: 
\begin{eqnarray}
&\mathcal{O}^{m_s=1/2}(\vec{p},t)=\mathcal{O}^{\mu=1}(\vec{p},t),\notag\\
&\mathcal{O}^{m_s=-1/2}(\vec{p},t)=\mathcal{O}^{\mu=2}(\vec{p},t),
\end{eqnarray}
where the $\mathcal{O}_{1,2}$ are the upper two components of Dirac four-spinor $\mathcal{O}_{\mu=1,...,4}$ in the Dirac basis. The positive-parity is captured by the upper two components of the spinor in the Dirac basis.

The  correlation function, with a definite three-momentum $\vec{p}$, for proton and $\Lambda$ are defined respectively as:
\begin{eqnarray}
    C_{p}(\vec{p},t)&=&\langle  \mathcal{O}_{p}(\vec{p},t)({\mathcal{O}}_{p}(\vec{p},0))^{\dagger} \rangle\notag\\
    C_{\Lambda}(\vec{p},t)&=&\langle  \mathcal{O}_{\Lambda}(\vec{p},t)({\mathcal{O}}_{\Lambda}(\vec{p},0))^{\dagger} \rangle.
\end{eqnarray}

\begin{figure}[h!]
\centering
\begin{subfigure}[b]{0.45\textwidth}
\includegraphics[width=1\textwidth]{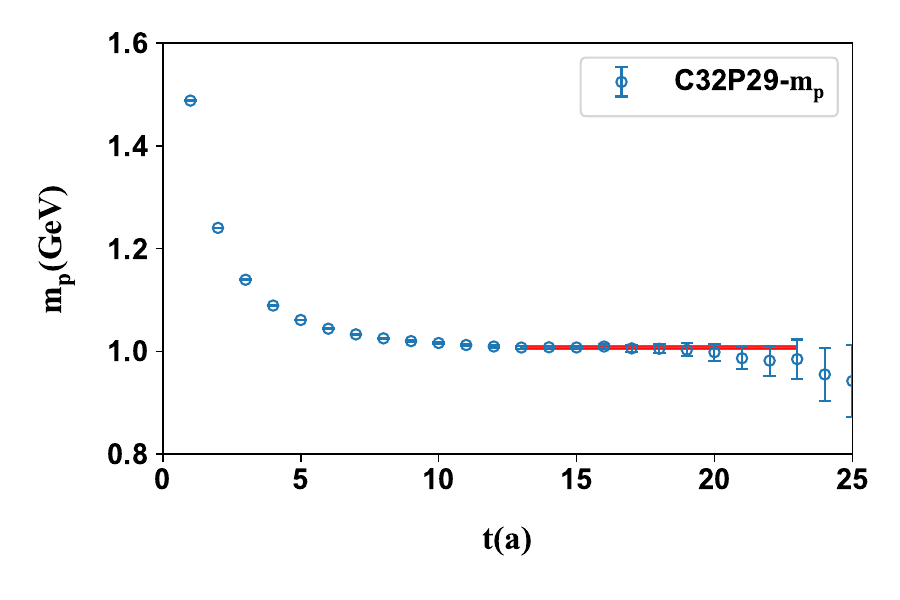} 
\end{subfigure}
\begin{subfigure}[b]{0.45\textwidth}
\includegraphics[width=1\textwidth]{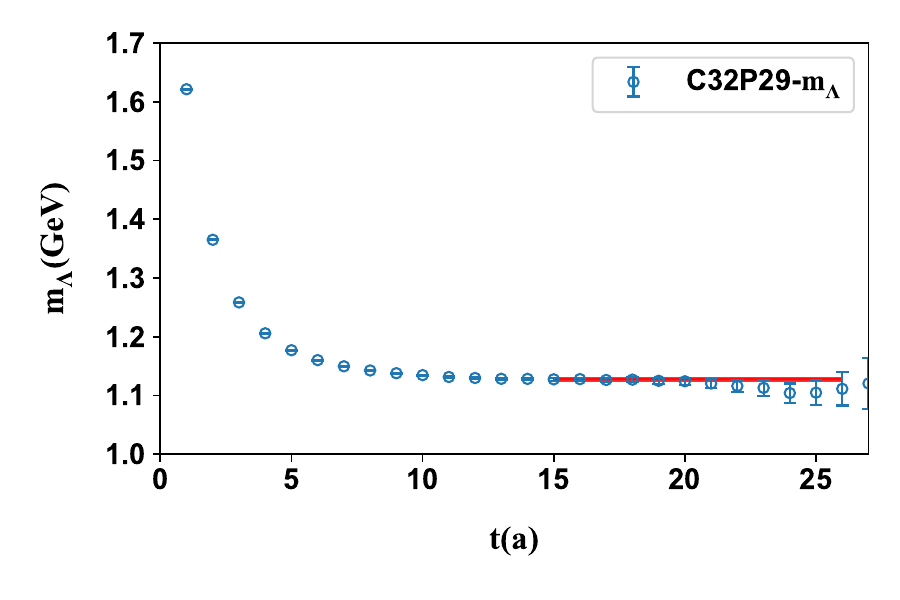} 
\end{subfigure}
\caption{The ground energies of the proton and $\Lambda$ on the ensemble C32P29. The horizontal bars indicate the fitted masses and the fit ranges.}
\label{fig: Effective-mass-of-single-particle}
\end{figure}

To examine the mass-energy dispersion relations of these two baryons and ensure the lattice discretization effects are well-controlled, we generate two-point correlation functions for the proton and \( \Lambda \) baryon at different momenta. From these correlation functions, the baryon energies \( E_{p}(\vec{p}) \) and \( E_{\Lambda}(\vec{p}) \) for various lattice momenta \( \vec{p} \) can be readily obtained. 
Fig.~\ref{fig:dispersion-C32P29} illustrates the dispersion relation, with the left panel showing the dispersion relation for the proton and the right for $\Lambda$ on the ensemble C32P29, where five momenta \( \vec{p} = [0,0,0], [0,0,1], [0,1,1], [1,1,1], [0,0,2] \frac{2\pi}{L} \) are chosen. The dispersion relation is then parametrized as follows:\begin{eqnarray}
  E^2=m^2+Z p^2.  
\label{eq:disper}
\end{eqnarray}
Here, \( Z \) is a parameter determined through fitting. As shown in Fig.~\ref{fig:dispersion-C32P29}, all lattice results are well-described by Eq.~\eqref{eq:disper}, yielding a reasonable $\chi^2/\text{d.o.f.}=0.2$ for proton and $\chi^2/\text{d.o.f.}=0.71$ for $\Lambda$. The values \( Z_{p} = 0.982(16) \) and \( Z_{\Lambda} = 0.985(11) \) are consistent with the square of the speed of light.

\begin{figure}[h]
\centering
\includegraphics[width=0.45\textwidth]{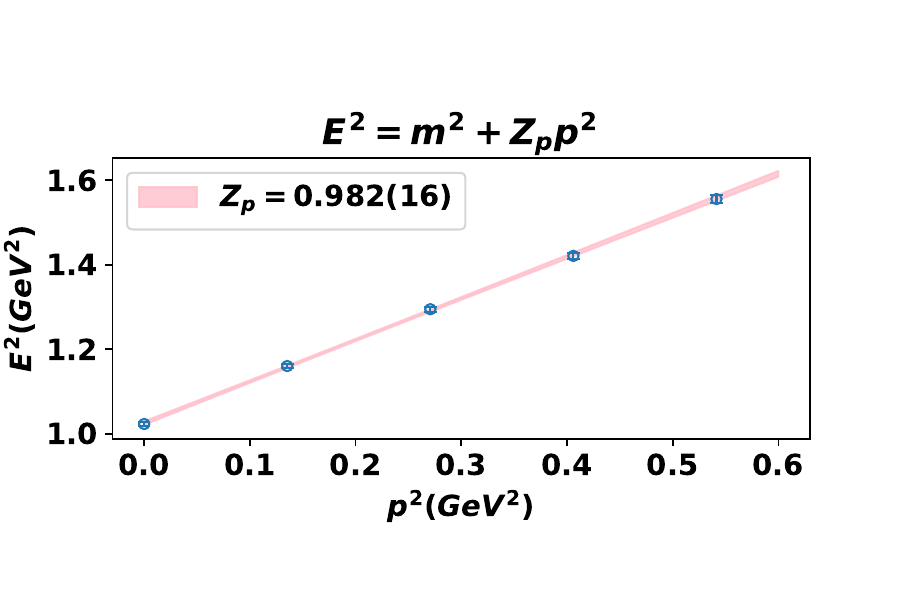} 
\includegraphics[width=0.45\textwidth]{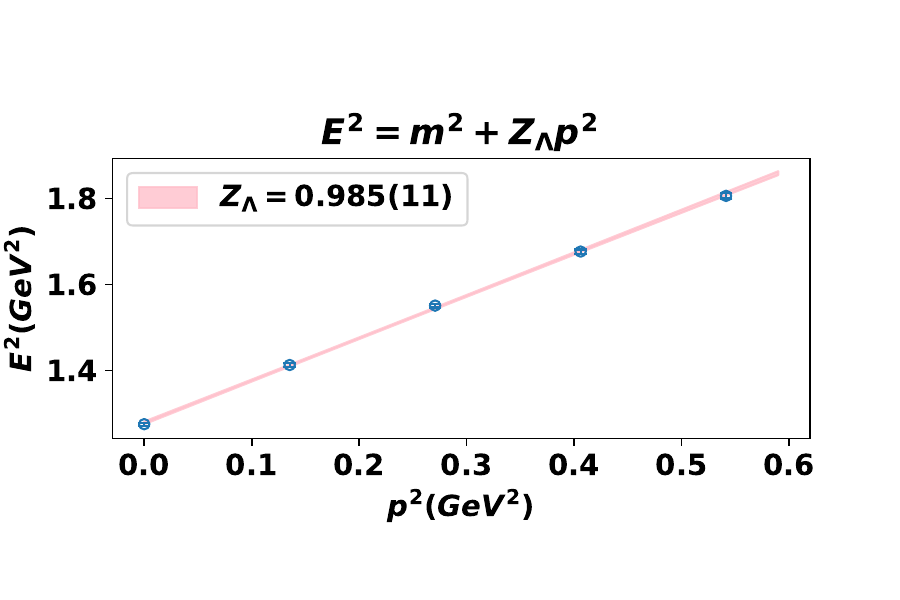}
\caption{Dispersion relation for proton (upper panel) and for $\Lambda$ on the C32P29 ensembles(lower panel). }
\label{fig:dispersion-C32P29}
\end{figure}

\section{Two-baryon operators and  correlation functions}

For two-baryon states, the operators must encode the system’s relevant symmetries, including relative spatial separation and total angular momentum, to accurately describe interparticle interactions. Constructing such two-particle operators further demands careful accounting of relevant internal quantum numbers. On the lattice, the continuous rotational symmetry group  SO(3)
is broken to the corresponding discrete point group. Here, we focus on the  $^1S_0$ and $^3S_1$ channels, which correspond to the irreps $A_1^+$ and  $T_1^+$ (respectively) in the center-of-mass frame, and utilize the publicly available \verb|OpTion| package~\cite{Yan:2025jlq} to perform group-theoretic projection.

The corresponding  correlation matrix are constructed as:
\begin{eqnarray}
     C_{ij}(t,t_s)=\langle \mathcal{O}_{i}(t)\mathcal{O}^{\dagger}_{j}(t_s) \rangle,
\end{eqnarray}
where $\mathcal{O}_i$ represents the two-particle operator defined in Eq.\eqref{eq:two-particle operator}, and $i$ and $j$ can take values 1 or 2.
The two-particle energies, which are substituted into  L\"uscher's formula, are extracted from this correlation matrix by solving the generalized eigenvalue problem (GEVP):
\begin{eqnarray}
    C(t)\cdot v_n(t,t_0)=\lambda_n(t,t_0)C(t_0)\cdot v_n(t,t_0),
\end{eqnarray}
with $n=1,2,...,N$ and $t>t_0$. The eigenvalues $\lambda_n(t,t_0)$ can be shown to behave like
\begin{eqnarray}
    \lambda_n\simeq e^{-E_n(t-t_0)}(1+C_1 e^{-\Delta E(t-t_0)}).
\label{eq:eigenvalue}
\end{eqnarray}
The finite-volume energy eigenvalues $E_n$ are extracted from the correlation matrix and mapped to the interacting momentum $q^2$, which serves as the input for Lüscher’s formula to determine scattering parameters. The reference time $t_0$ is optimized to suppress excited-state contamination while maintaining a sufficient signal-to-noise ratio.

\subsection{Two-particle operators 
Constructing in the center-of-mass frame}
In this work, the publicly available package \verb|OpTion|~\cite{Yan:2025jlq} is utilized to construct  the group theory projection. The scattering analysis is restricted to the $A_1^+$ ($^1S_0$) and $T_1^+$ ($^3S_1$) irreps of the cubic group ($O_h$) at rest: 
\begin{eqnarray}
A_1^+:&\notag\\
\mathcal{O}^1_{A_1^+}&=&\mathcal{O}^p_{\mu=\frac{1}{2}}(\vec{0},t)\mathcal{O}^{\Lambda}_{\mu=-\frac{1}{2}}(\vec{0},t)-\mathcal{O}^p_{\mu=-\frac{1}{2}}(\vec{0},t)\mathcal{O}^{\Lambda}_{\mu=\frac{1}{2}}(\vec{0},t)\notag\\
\mathcal{O}^2_{A_1^+}&=&\mathcal{O}^p_{\mu=\frac{1}{2}}(\vec{e}_x,t)\mathcal{O}^{\Lambda}_{\mu=-\frac{1}{2}}(-\vec{e}_x,t)-\mathcal{O}^p_{\mu=-\frac{1}{2}}(\vec{e}_x,t)\mathcal{O}^{\Lambda}_{\mu=\frac{1}{2}}(-\vec{e}_x,t)\notag\\
&+&\mathcal{O}^p_{\mu=\frac{1}{2}}(-\vec{e}_x,t)\mathcal{O}^{\Lambda}_{\mu=-\frac{1}{2}}(\vec{e}_x,t)-\mathcal{O}^p_{\mu=-\frac{1}{2}}(-\vec{e}_x,t)\mathcal{O}^{\Lambda}_{\mu=\frac{1}{2}}(\vec{e}_x,t)\notag\\
&+&\mathcal{O}^p_{\mu=\frac{1}{2}}(\vec{e}_y,t)\mathcal{O}^{\Lambda}_{\mu=-\frac{1}{2}}(-\vec{e}_y,t)-\mathcal{O}^p_{\mu=-\frac{1}{2}}(\vec{e}_y,t)\mathcal{O}^{\Lambda}_{\mu=\frac{1}{2}}(-\vec{e}_y,t)\notag\\
&+&\mathcal{O}^p_{\mu=\frac{1}{2}}(-\vec{e}_y,t)\mathcal{O}^{\Lambda}_{\mu=-\frac{1}{2}}(\vec{e}_y,t)-\mathcal{O}^p_{\mu=-\frac{1}{2}}(-\vec{e}_y,t)\mathcal{O}^{\Lambda}_{\mu=\frac{1}{2}}(\vec{e}_y,t)\notag\\
&+&\mathcal{O}^p_{\mu=\frac{1}{2}}(\vec{e}_z,t)\mathcal{O}^{\Lambda}_{\mu=-\frac{1}{2}}(-\vec{e}_z,t)-\mathcal{O}^p_{\mu=-\frac{1}{2}}(\vec{e}_z,t)\mathcal{O}^{\Lambda}_{\mu=\frac{1}{2}}(-\vec{e}_z,t)\notag\\
&+&\mathcal{O}^p_{\mu=\frac{1}{2}}(-\vec{e}_z,t)\mathcal{O}^{\Lambda}_{\mu=-\frac{1}{2}}(\vec{e}_z,t)-\mathcal{O}^p_{\mu=-\frac{1}{2}}(-\vec{e}_z,t)\mathcal{O}^{\Lambda}_{\mu=\frac{1}{2}}(\vec{e}_z,t),\\
T_1^+:&\notag\\
\mathcal{O}^1_{T_1^+}&=&\mathcal{O}^p_{\mu=-\frac{1}{2}}(\vec{0},t)\mathcal{O}^{\Lambda}_{\mu=-\frac{1}{2}}(\vec{0},t)-\mathcal{O}^p_{\mu=\frac{1}{2}}(\vec{0},t)\mathcal{O}^{\Lambda}_{\mu=\frac{1}{2}}(\vec{0},t)\notag\\
\mathcal{O}^2_{T_1^+}&=&\mathcal{O}^p_{\mu=-\frac{1}{2}}(\vec{e}_x,t)\mathcal{O}^{\Lambda}_{\mu=-\frac{1}{2}}(-\vec{e}_x,t)-\mathcal{O}^p_{\mu=\frac{1}{2}}(\vec{e}_x,t)\mathcal{O}^{\Lambda}_{\mu=\frac{1}{2}}(-\vec{e}_x,t)\notag\\
&+&\mathcal{O}^p_{\mu=-\frac{1}{2}}(-\vec{e}_x,t)\mathcal{O}^{\Lambda}_{\mu=-\frac{1}{2}}(\vec{e}_x,t)-\mathcal{O}^p_{\mu=\frac{1}{2}}(-\vec{e}_x,t)\mathcal{O}^{\Lambda}_{\mu=\frac{1}{2}}(\vec{e}_x,t)\notag\\
&+&\mathcal{O}^p_{\mu=-\frac{1}{2}}(\vec{e}_y,t)\mathcal{O}^{\Lambda}_{\mu=-\frac{1}{2}}(-\vec{e}_y,t)-\mathcal{O}^p_{\mu=\frac{1}{2}}(\vec{e}_y,t)\mathcal{O}^{\Lambda}_{\mu=\frac{1}{2}}(-\vec{e}_y,t)\notag\\
&+&\mathcal{O}^p_{\mu=-\frac{1}{2}}(-\vec{e}_y,t)\mathcal{O}^{\Lambda}_{\mu=-\frac{1}{2}}(\vec{e}_y,t)-\mathcal{O}^p_{\mu=\frac{1}{2}}(-\vec{e}_y,t)\mathcal{O}^{\Lambda}_{\mu=\frac{1}{2}}(\vec{e}_y,t)\notag\\
&+&\mathcal{O}^p_{\mu=-\frac{1}{2}}(\vec{e}_z,t)\mathcal{O}^{\Lambda}_{\mu=-\frac{1}{2}}(-\vec{e}_z,t)-\mathcal{O}^p_{\mu=\frac{1}{2}}(\vec{e}_z,t)\mathcal{O}^{\Lambda}_{\mu=\frac{1}{2}}(-\vec{e}_z,t)\notag\\
&+&\mathcal{O}^p_{\mu=-\frac{1}{2}}(-\vec{e}_z,t)\mathcal{O}^{\Lambda}_{\mu=-\frac{1}{2}}(\vec{e}_z,t)-\mathcal{O}^p_{\mu=\frac{1}{2}}(-\vec{e}_z,t)\mathcal{O}^{\Lambda}_{\mu=\frac{1}{2}}(\vec{e}_z,t). 
\label{eq:two-particle operator}
\end{eqnarray}

\subsection{Two particles scattering in
moving frames in a finite volume}
There are two particles with total three-momentum $\mathbf P$ in a box, and the total momentum satisfies
\begin{eqnarray}
    \mathbf P=\mathbf p_1 +\mathbf p_2=\frac{2\pi}{L}\mathbf{d}, \mathbf{d}\in Z^3.
\end{eqnarray}
Then the energy $E^{*}$ is 
\begin{eqnarray}
E^{* 2}= E^2_{MF}-\mathbf{P}^2.
\end{eqnarray}
In the moving frame, the center-of-mass $E^{*}$ is moving with a velocity of $v=\mathbf{P}/E_{MF}$ and the Lorentz factor $\gamma = (1 - v^2)^{-1/2}$.
To get more eigen-energies near the threshold, we implement the moving frame $d=e_3$. Considering the two particles with different masses, a 2×2 correlation function matrix can be constructed, simulated, and diagonalized to yield the eigen-energies.We focus on the $^1S_0 $ and $^3S_1$ channels, which correspond to the irreps  $A_1$ and $E_2$ in this moving frame, respectively: 
\begin{eqnarray}
A_1:&\notag\\
\mathcal{O}^1_{A_1}&=&\mathcal{O}^p_{\mu=\frac{1}{2}}(\vec{0},t)\mathcal{O}^{\Lambda}_{\mu=-\frac{1}{2}}(\vec{e_z},t)-\mathcal{O}^p_{\mu=-\frac{1}{2}}(\vec{0},t)\mathcal{O}^{\Lambda}_{\mu=\frac{1}{2}}(\vec{e_z},t),\notag\\
\mathcal{O}^2_{A_1}&=&\mathcal{O}^p_{\mu=\frac{1}{2}}(\vec{e_z},t)\mathcal{O}^{\Lambda}_{\mu=-\frac{1}{2}}(\vec{0},t)-\mathcal{O}^p_{\mu=-\frac{1}{2}}(\vec{e_z},t)\mathcal{O}^{\Lambda}_{\mu=\frac{1}{2}}(\vec{0},t),\notag\\
E_2:&\notag\\
\mathcal{O}^1_{E_2}&=&\mathcal{O}^p_{\mu=-\frac{1}{2}}(\vec{0},t)\mathcal{O}^{\Lambda}_{\mu=-\frac{1}{2}}(\vec{e_z},t)+\mathcal{O}^p_{\mu=\frac{1}{2}}(\vec{0},t)\mathcal{O}^{\Lambda}_{\mu=\frac{1}{2}}(\vec{e_z},t),\notag\\
\mathcal{O}^2_{E_2}&=&\mathcal{O}^p_{\mu=-\frac{1}{2}}(\vec{e_z},t)\mathcal{O}^{\Lambda}_{\mu=-\frac{1}{2}}(\vec{0},t)+\mathcal{O}^p_{\mu=\frac{1}{2}}(\vec{e_z},t)\mathcal{O}^{\Lambda}_{\mu=\frac{1}{2}}(\vec{0},t). 
\label{eq:moving-frame-operators}
\end{eqnarray}
Then the scattering phase shifts can be gained from the total energy of the two-particle system enveloped in a cubic torus by L\"uscher technique ~\cite{Luscher:1986pf, Luscher:1990ux, Luscher:1990ck, Rummukainen:1995vs}
\begin{eqnarray}
q \cot{\delta_0(q)} = \frac{1}{\gamma \pi^{3/2}} \mathcal{Z}^d_{00}(1; q^2).
\end{eqnarray}
The zeta function $\mathcal{Z}^d_{00}(1; q^2)$ is computed  using the expression~\cite{NPLQCD:2011htk}:
\begin{eqnarray}
\begin{aligned} 
\mathcal{Z}_{L M}^{d}\left(1 ; q^{ 2}\right) & =\sum_{\mathbf{r}\in P_d} \frac{e^{-\Lambda\left(|\mathbf{r}|^2-q^{2}\right)}}{|\mathbf{r}|^2-q^{2}}|\mathbf{r}|^L Y_{L M}\left(\Omega_{\mathbf{r}}\right) \\ & +\delta_{L, 0} Y_{00} \gamma \pi^{3 / 2}\left[2 q^{2} \int_0^{\Lambda} d t \frac{e^{t q^{2}}}{\sqrt{t}}-\frac{2}{\sqrt{\Lambda}} e^{\Lambda q^{2}}\right] \\ & +\gamma \sum_{\mathbf{w} \neq \mathbf{0}} e^{-i \pi \mathbf{w} \cdot \mathbf{Ad}}|\hat{\gamma} \mathbf{w}|^L Y_{L M}\left(\Omega_{\hat{\gamma} \mathbf{w}}\right) \int_0^{\Lambda} d t\left(\frac{\pi}{t}\right)^{3 / 2+L} e^{t q^{2}} e^{-\frac{\pi^2|\hat{\gamma} \mathbf{w}|^2}{t}}, 
\end{aligned}
\end{eqnarray}
where $Y_{LM}$ are spherical harmonics. The factor A accounts for the mass difference of the two scattering particles and is defined as $A = 1+\frac{m_1^2 - m_2^2}{E^{*2}}$. The sum of  $\mathbf{r}$ runs over the set $P_d$ defined by $P_d=\,\{\mathbf{r}\,|\,\mathbf{r}=\hat{\gamma}^{-1}(\mathbf{n}-\frac{1}{2} A \mathbf{d}),\mathbf{n}\in Z^3\}$. The value of above expression is independent of the choice of the cutoff parameter $\Lambda$. In this work, we set $\Lambda=1$. 
 
\section{The two-particle spectrum}

We show the results for the two-particle spectrum  in Fig.~\ref{fig:Effective-deltaE}, \ref{fig:Effective-deltaE-moving-frame}, \ref{fig:Effective-deltaE-T1}, \ref{fig:Effective-deltaE-moving-frame-E2}.

\begin{figure}[h!]
\centering
\includegraphics[width=0.45\textwidth]{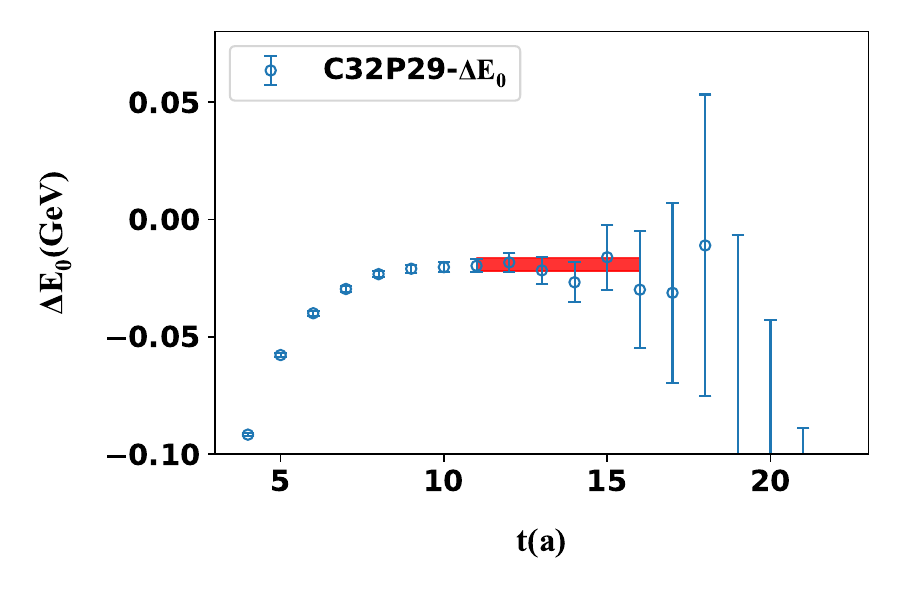}
\caption{Effective energy splitting $\Delta E_\alpha=\log \frac{\mathcal{R}_{\alpha}(t+1,t_0)}{\mathcal{R}_{\alpha}(t,t_0)}$ for the irrep $A^{+}_1$ on the ensemble C32P29 in the center-of-mass frame. The horizontal bars indicate the fitted $\Delta E_\alpha$ and the fit ranges.} 
\label{fig:Effective-deltaE}
\end{figure} 
 
\begin{figure}[h!]
\centering
\begin{subfigure}[b]{0.45\textwidth}
\includegraphics[width=1\textwidth]{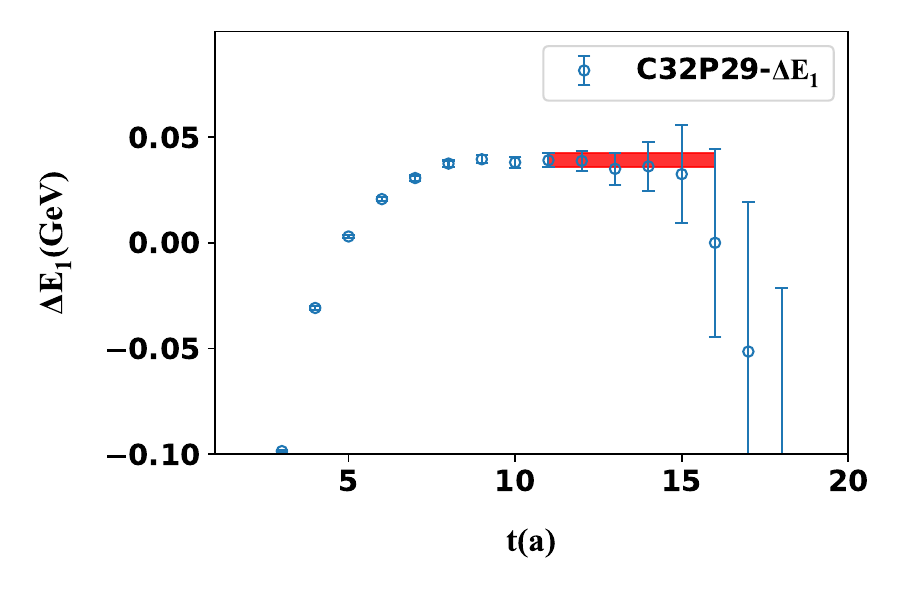} 
\end{subfigure}
\begin{subfigure}[b]{0.45\textwidth}
\includegraphics[width=1\textwidth]{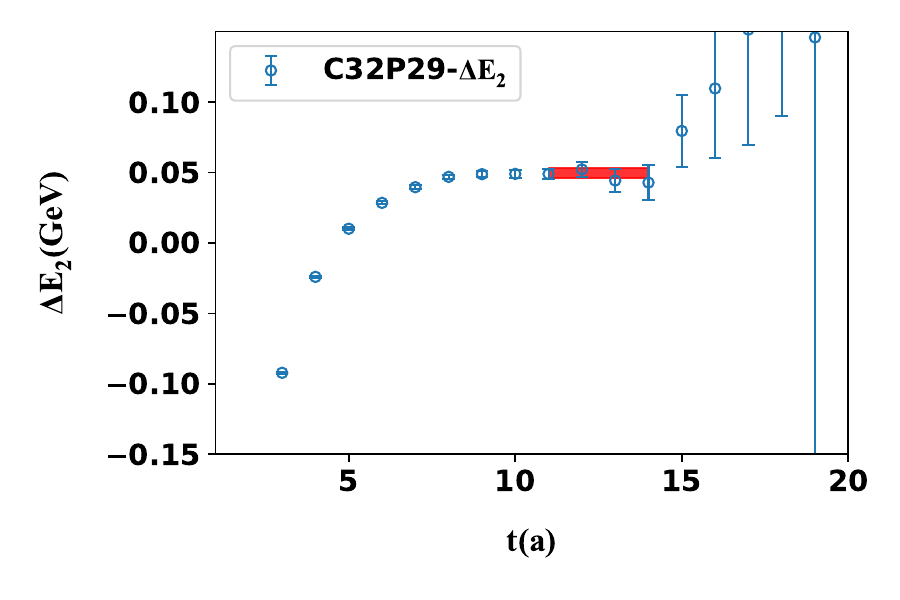} 
\end{subfigure}
\caption{The two Effective energy splittings for the irrep $A_1$ on the ensemble C32P29 in the moving frame $d=e_3$. The horizontal bars indicate the fitted $\Delta E_\alpha$ and the fit ranges.} 
\label{fig:Effective-deltaE-moving-frame}
\end{figure}

\begin{figure}[h!]
\centering
\includegraphics[width=0.45\textwidth]{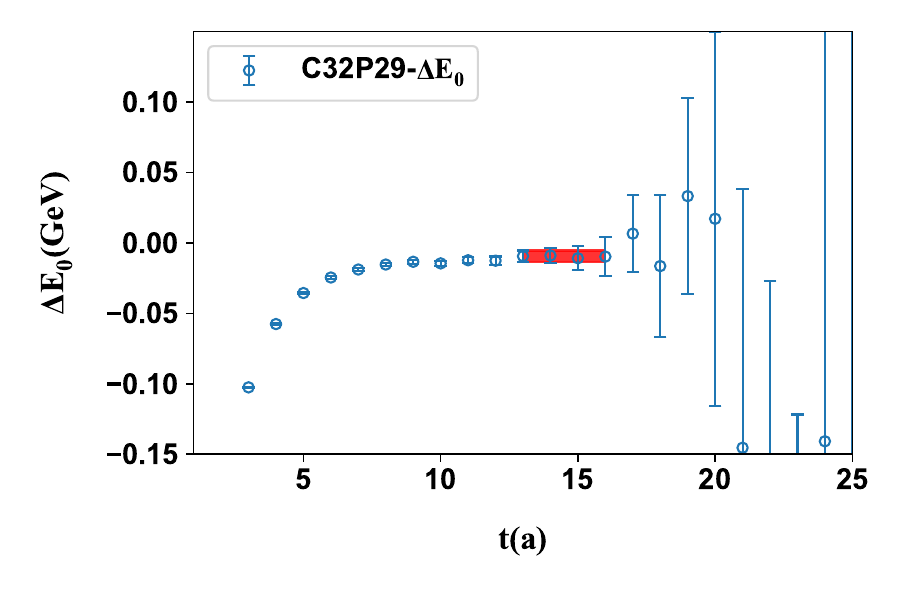}
\caption{Effective energy splitting $\Delta E_\alpha=\log \frac{\mathcal{R}_{\alpha}(t+1,t_0)}{\mathcal{R}_{\alpha}(t,t_0)}$ for the irrep $T^{+}_1$ on the ensemble C32P29 in the center-of-mass frame. The horizontal bars indicate the fitted $\Delta E_\alpha$ and the fit ranges.} 
\label{fig:Effective-deltaE-T1}
\end{figure} 
 
\begin{figure}[h!]
\centering
\begin{subfigure}[b]{0.45\textwidth}
\includegraphics[width=1\textwidth]{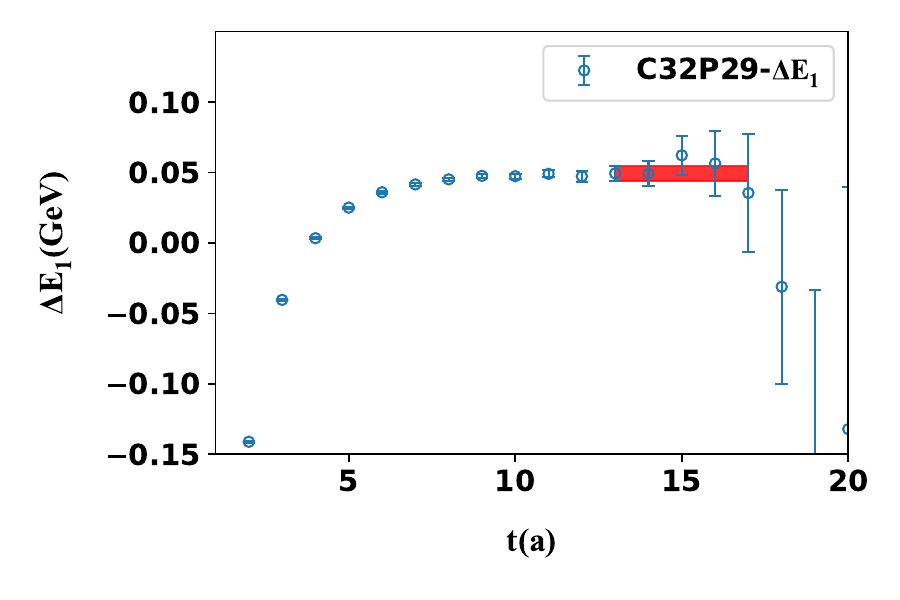} 
\end{subfigure}
\caption{The Effective energy splitting for the irrep $E_2$ on the ensemble C32P29 in the moving frame $d=e_3$. The horizontal bars indicate the fitted $\Delta E_\alpha$ and the fit ranges.} 
\label{fig:Effective-deltaE-moving-frame-E2}
\end{figure}

\section{ERE expansion diagrams of other ensembles}

We show the results for ERE expansion diagrams of other ensembles in Fig.~\ref{fig:ERE_fit_others-A1}, and Fig.~\ref{fig:T1_ERE_fit_others}. In Fig.~\ref{fig:ERE_fit_others-A1}, 
results for the parameter fits according to Eq.~\eqref{eq:ERE} for the channel $^1S_0$ are given. The points are from lattice data from the coresspending ensemble. Both horizontal and vertical coordinates are expressed in lattice units. The gray band corresponds to statistical error. Intersection of the $ik$ curve (black) and the gray band indicates the allowed pole range. Results for the  $^3S_1$ channel are given in Fig.~\ref{fig:T1_ERE_fit_others}. 

%%%%%%%%%%%%%%%%%%%%%%%%%%%%%%%
\begin{figure}[h!]
\centering
\begin{subfigure}[b]{0.45\textwidth}
\includegraphics[width=1\textwidth]{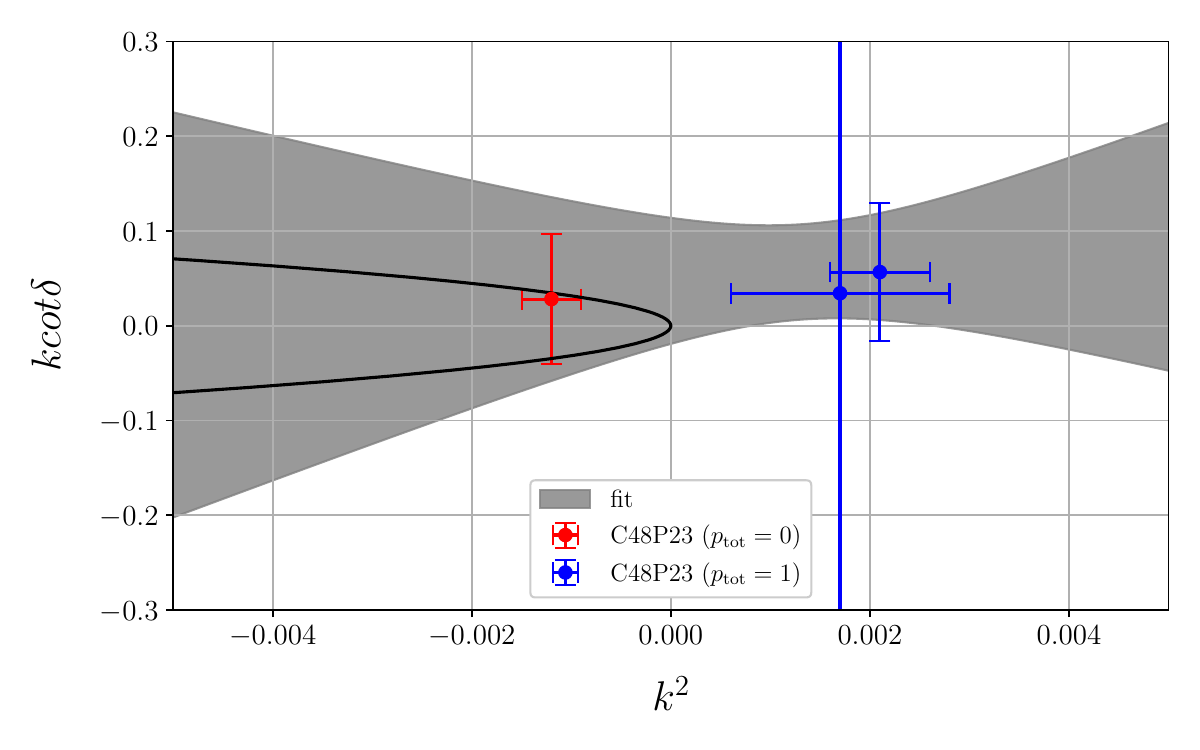} 
\caption{C48P23}
\end{subfigure}
\begin{subfigure}[b]{0.45\textwidth}
\includegraphics[width=1\textwidth]{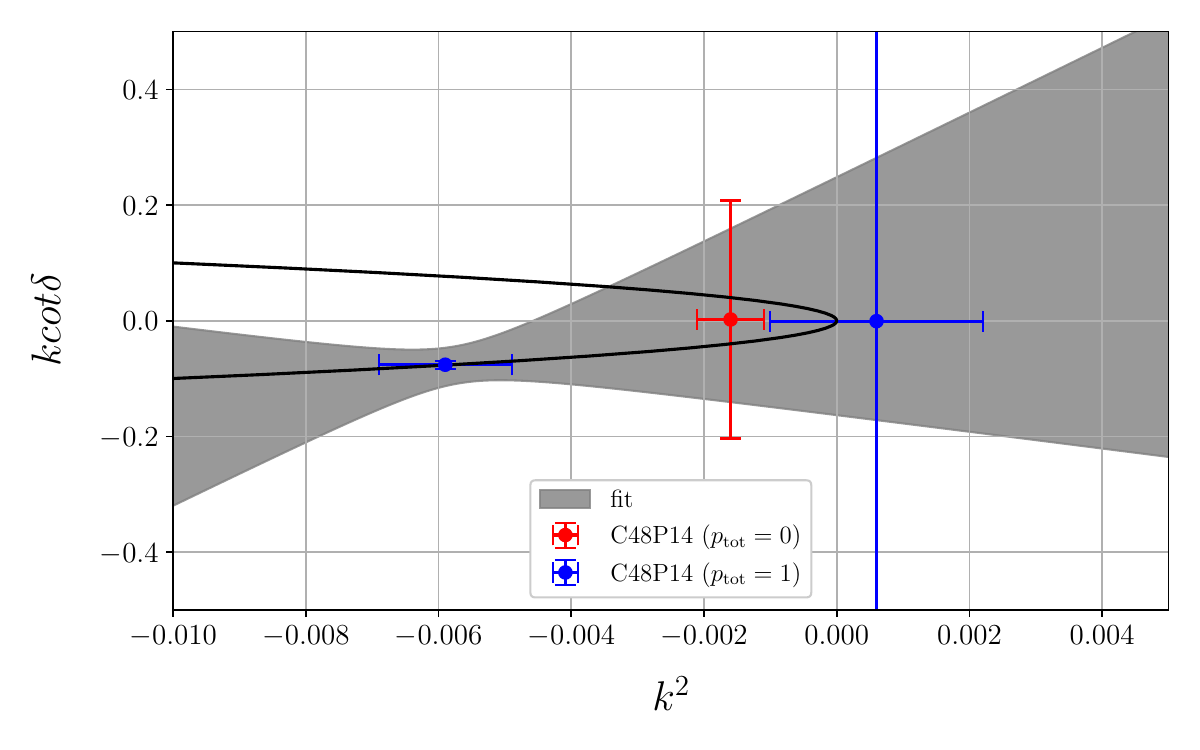} 
\caption{C48P14}
\end{subfigure}
\begin{subfigure}[b]{0.45\textwidth}
\includegraphics[width=1\textwidth]{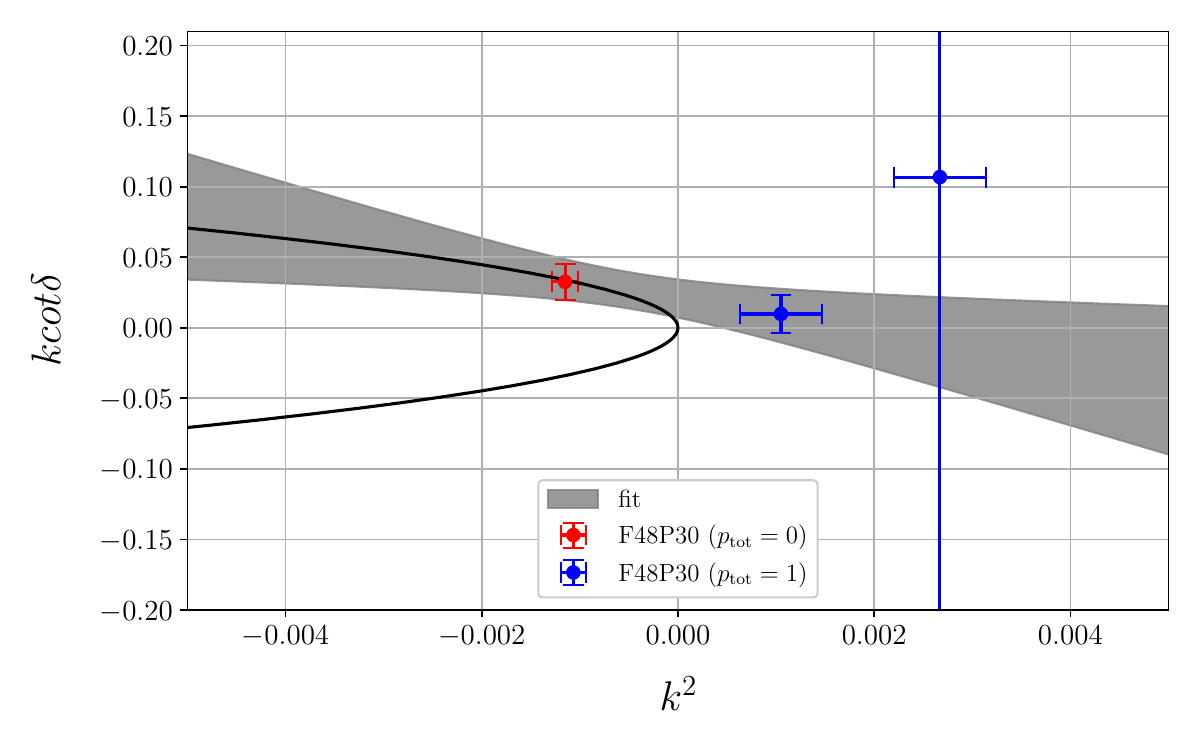} 
\caption{F48P30}
\end{subfigure}
\begin{subfigure}[b]{0.45\textwidth}
\includegraphics[width=1\textwidth]{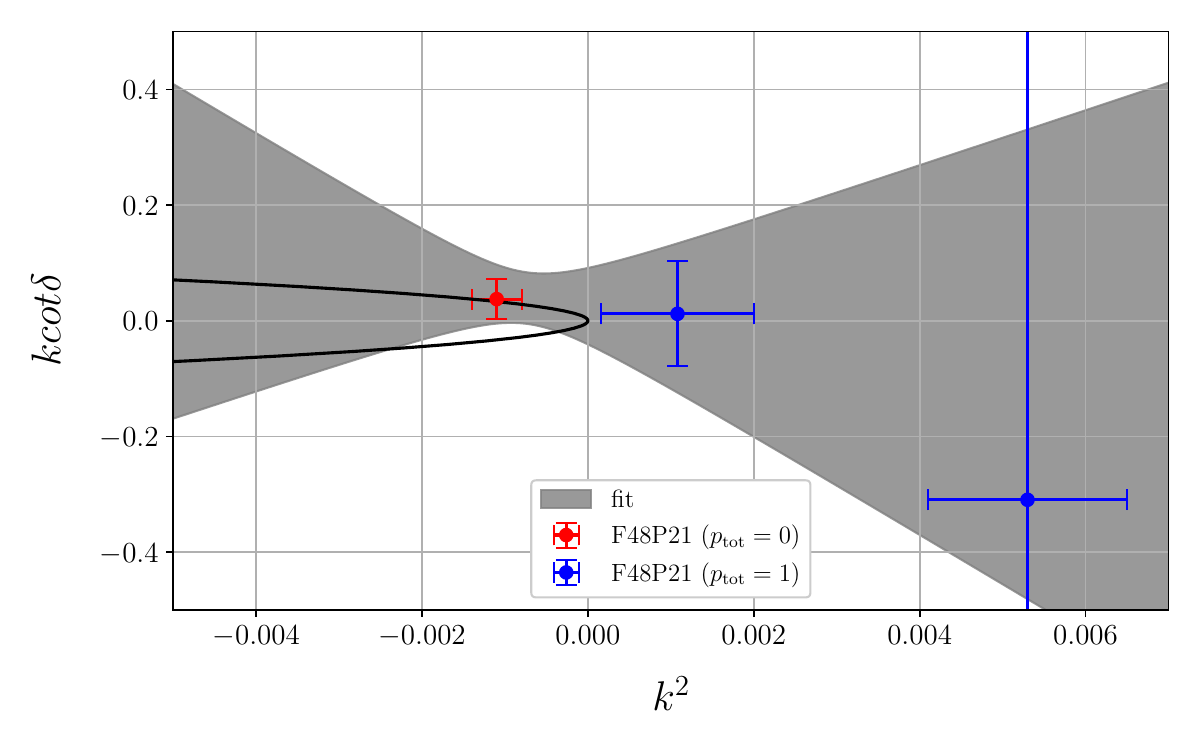} 
\caption{F48P21}
\end{subfigure}
\begin{subfigure}[b]{0.45\textwidth}
\includegraphics[width=1\textwidth]{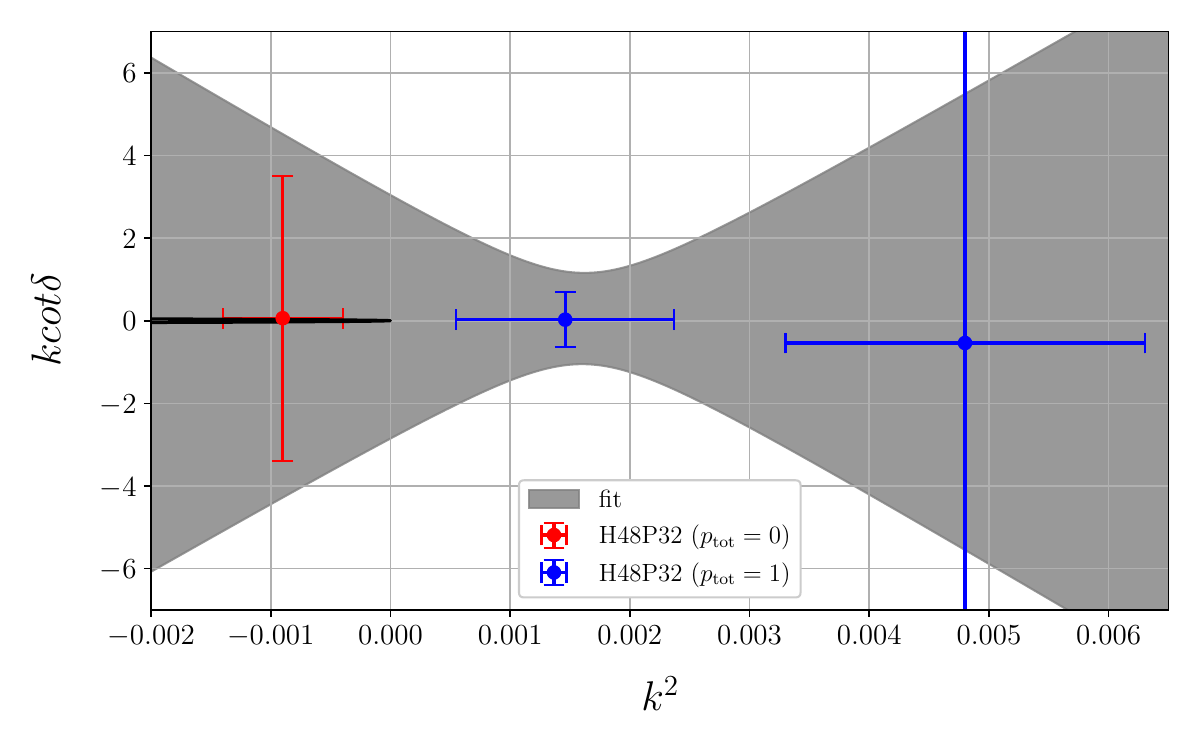} 
\caption{H48P32}
\end{subfigure}
\caption{Results for the parameter fits according to Eq.~\eqref{eq:ERE} for the channel $^1S_0$. The points are from lattice data from the coresspending ensemble. Both horizontal and vertical coordinates are expressed in lattice units. The gray band corresponds to statistical error. Intersection of the $ik$ curve (black) and the gray band indicates the allowed pole range. } 
\label{fig:ERE_fit_others-A1}
\end{figure}

\begin{figure}[h!]
\centering
\begin{subfigure}[b]{0.45\textwidth}
\includegraphics[width=1\textwidth]{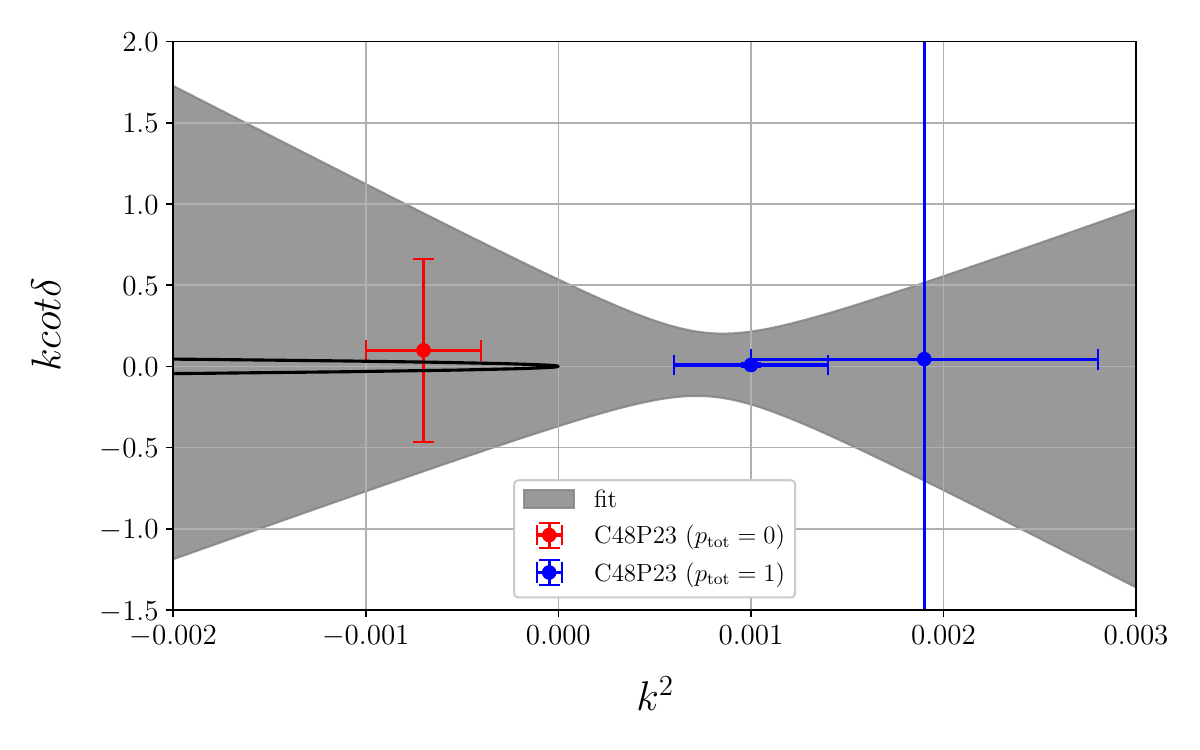} 
\caption{C48P23}
\end{subfigure}
\begin{subfigure}[b]{0.45\textwidth}
\includegraphics[width=1\textwidth]{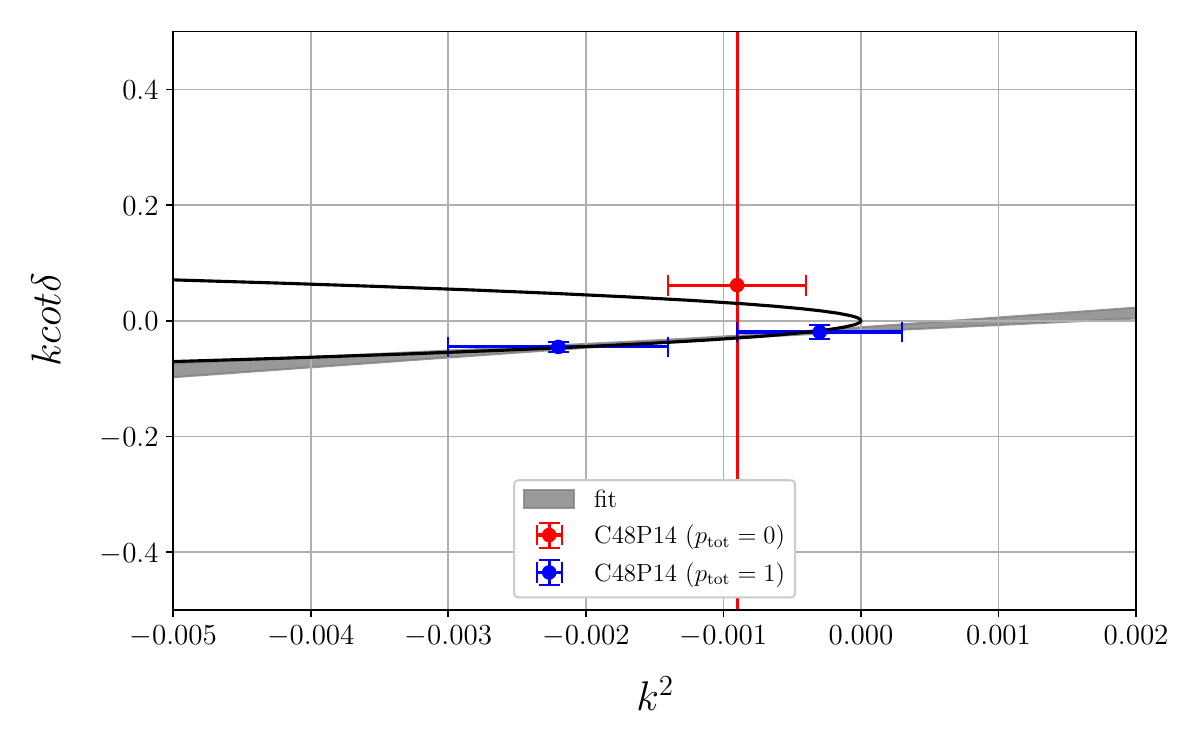}
\caption{C48P14}
\end{subfigure}
\begin{subfigure}[b]{0.45\textwidth}
\includegraphics[width=1\textwidth]{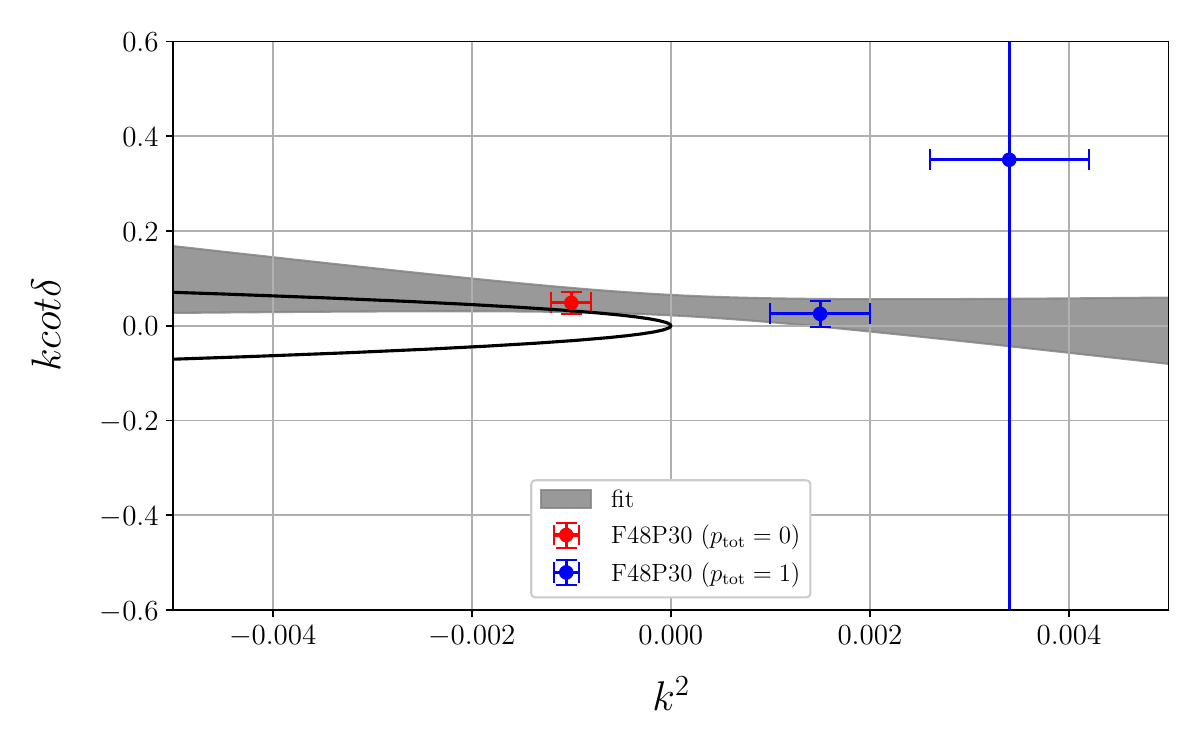}
\caption{F48P30}
\end{subfigure}
\begin{subfigure}[b]{0.45\textwidth}
\includegraphics[width=1\textwidth]{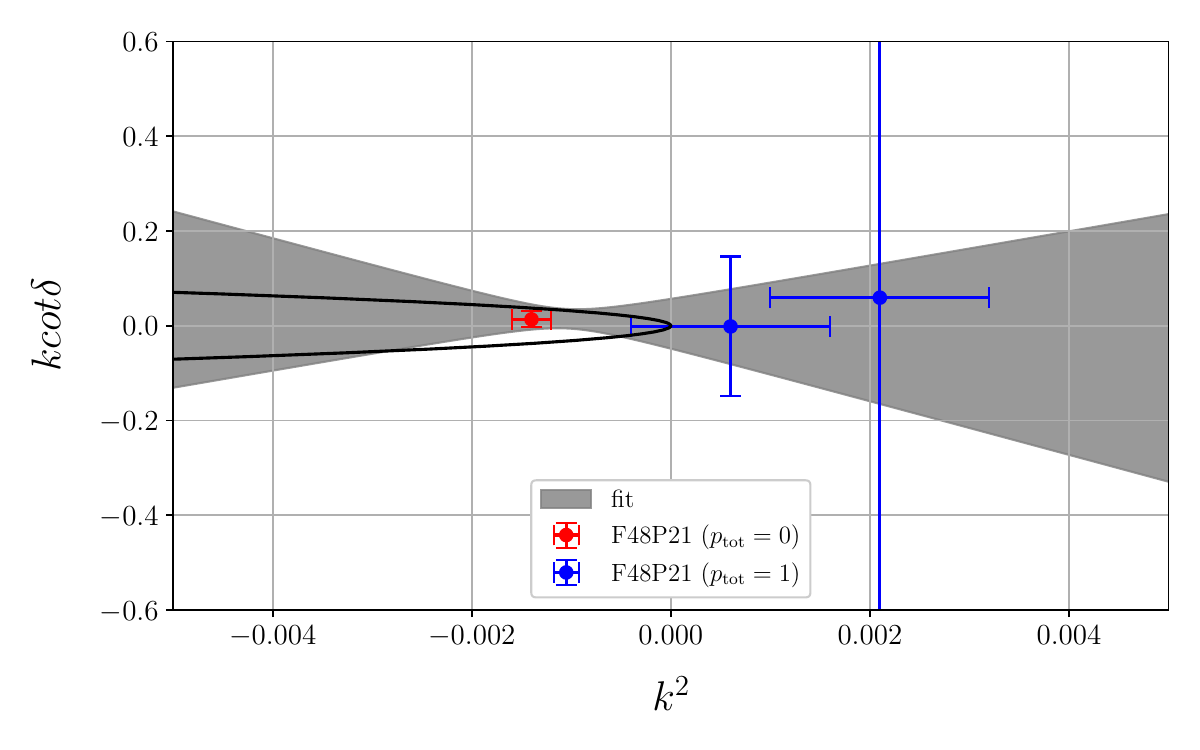}
\caption{F48P21}
\end{subfigure}
\begin{subfigure}[b]{0.45\textwidth}
\includegraphics[width=1\textwidth]{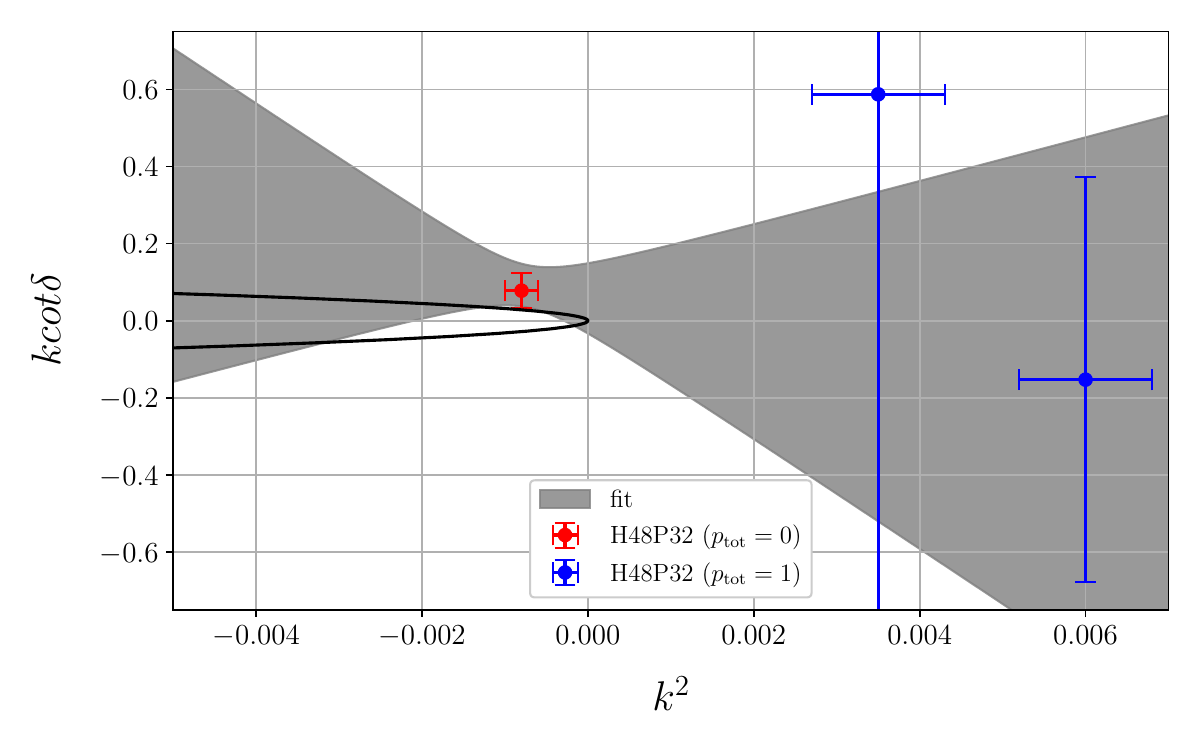}
\caption{H48P32}
\end{subfigure}
\caption{Results for the parameter fits according to Eq.~\eqref{eq:ERE} for the channel $^3S_1$. The gray band corresponds to statistical error.} 
\label{fig:T1_ERE_fit_others}
\end{figure}
\end{widetext}

\end{appendix}

\end{document}